\documentclass[]{aa}  

\usepackage{graphicx}
\usepackage{txfonts}
\usepackage{subcaption}         
\usepackage{lscape}            
\usepackage{placeins}    

\newcommand{\ocen}{$\omega$ Cen}

\begin{document}

   \title{Studying stellar populations in Omega Centauri with phylogenetics}

   \author{P. Jofré
          \inst{1,2}
        \and  C. Aguilera-Gómez\inst{2,3}
        \and  P. Villarreal\inst{4,5,6} 
        \and F. A. Cubillos\inst{4,5}
        \and P. Das\inst{7}
        \and X. Hua\inst{8}
        \and R. Yates\inst{9}
        \and P. Silva\inst{2,3}
        \and S. Vitali\inst{1}
        \and T. Peña\inst{4,5}
        \and T. Signor\inst{1,10}
        \and K. Walsen\inst{1,2,11}
        \and P. Tissera\inst{2,3,12}
        \and A. Rojas-Arriagada\inst{2,13,14,15}
        \and E. Johnston\inst{1,2}
        \and G. Gilmore\inst{16,17}
        \and R. Foley\inst{18}
          }

   \institute{Instituto de Estudios Astrofísicos, Facultad de Ingenier\'ia y Ciencias, Universidad Diego Portales, Ejército Libertador 441, Santiago, Chile \email{paula.jofre@mail.udp.cl}
   \and 
    Millennium Nucleus ERIS
    \and 
    Instituto de Astrof\'isica, Pontificia Universidad Cat\'olica de Chile, Av. Vicu\~na Mackenna 4860, 782-0436 Macul, Santiago, Chile
    \and
    Millennium Institute for Integrative Biology (iBio), Santiago, Chile
    \and 
    Centro de Gen\'omica y Bioinform\'atica, Facultad de Ciencias, Ingenier\'ia y Tecnolog\'ia, Universidad Mayor, Santiago, Chile
\and Departamento de Biolog\'ia, Facultad de Qu\'imica y Biolog\'ia, Universidad de Santiago de \and School of Mathematics and Physics, University of Surrey, Guildford, Surrey, GU2 7XH, UK
\and Mathematical Sciences Institute, Australian National University, Canberra ACT 2601, Australia
\and Centre for Astrophysics Research, University of Hertfordshire, Hatfield, AL10 9AB, UK
\and Inria Chile Research Center, Av. Apoquindo 2827, piso 12, Las Condes, Santiago, Chile
\and Departamento de Ingenier\'ia Matem\'atica, Facultad de Ciencias F\'isicas y Matem\'aticas, Universidad de Chile, Av. Beauchef 851, 8370458 Santiago, Chile
\and Centro de Astroingeniería, Pontificia Universidad Católica de Santiago, Av. Vicuña Makenna 4860, Santiago, Chile
\and Departamento de F\'isica, Universidad de Santiago de Chile, Av. Victor Jara 3659, Santiago, Chile
\and Millenium Institute of Astrophysics (MAS), Av. Vicuña Mackenna 4860, 82-0436 Macul, Santiago, Chile
\and Center for Interdisciplinary Research in Astrophysics and Space Exploration (CIRAS), Universidad de Santiago de Chile, Santiago, Chile
\and Institute of Astronomy, University of Cambridge, Madingley Road,  CB3 OHA,  Cambridge, UK
\and Institute of Astrophysics, FORTH Cert, N. Plastira 100, GR-70013 Vassilika Vouton, Crete , Greece
\and Leverhulme Centre for Human Evolutionary Studies, Department of Archaeology, University of Cambridge, Fitzwilliam Street, Cambridge CB2 1QH, UK
   }

   \date{Received September 15, 1996; accepted March 16, 1997}

  \abstract
  {The nature and formation history of our Galaxy's largest and most enigmatic stellar cluster,  known as Omega Centauri (\ocen) remains debated. }
  {Here, we offer a novel approach to disentangling the complex stellar populations within \ocen\ based on phylogenetics methodologies from evolutionary biology.}
  {These include the Gaussian mixture model and neighbor-joining clustering algorithms applied to a set of chemical abundances of \ocen\  stellar members. 
  Instead of using the classical approach in astronomy of grouping them into separate populations, we focused on how the stars are related to each other.} {We could identify stars that likely formed in globular clusters versus those originating from prolonged  
in-situ star formation and how these stars interconnect.}
{Our analysis supports the hypothesis that \ocen\ might be a nuclear star cluster of a galaxy accreted by the Milky Way with a mass of about $10^9 \mathrm{M}_\odot$. Furthermore, we revealed the existence of a previously unidentified in-situ stellar population with a distinct chemical pattern unlike any known population found in the Milky Way to date. Our analysis of \ocen\ is an example of the success of cross-disciplinary research and shows the vast potential of applying evolutionary biology tools to astronomical datasets, opening new avenues for understanding the chemical evolution of complex stellar systems.}


   \maketitle
%

\section{Introduction}

Omega Centauri (\ocen) is the largest nearby stellar cluster, but its origin remains unresolved. Among such stellar clusters, globular clusters (GCs) are very old and dense, and tend to live in the halos of galaxies. While GCs were once expected to be be composed of a simple coeval stellar population with homogeneous chemical compositions, \ocen, like many other massive globular star clusters, hosts multiple stellar populations covering a wide range of iron abundances. A possible hypothesis is that \ocen\ might be a nuclear star cluster (NSC) that originated at the center of a galaxy accreted by the Milky Way \citep{BekkiFreeman03, AlvarezGaray24}, making \ocen\ a remnant of this galaxy.

Understanding how NSCs form and evolve is an active field of research in astrophysics \citep{Neumayer20}. These clusters form in the centers of galaxies, where the environment is extremely dense, packed with stars, GCs, and gas. Critically, an NSC is at the base of a very much deeper gravitational potential well than that of an isolated star cluster-forming molecular cloud. Thus much more complex, and longer duration evolutionary sequences will be encoded. Depending on the mass of the progenitor galaxy, NSCs are believed to be assembled via accretion of GCs or in-situ star formation fuelled by infalling gas \citep{Tremaine75, milos04, Fahrion21}. Nonetheless, neither the relative contribution of either channel nor the dependence on host galaxy properties is fully understood \citep{Fahrion21}. Disentangling the various evolutionary pathways can shed light on the assembly mechanisms of NSCs. In this context, \ocen, as an NSC candidate, provides a unique opportunity to study resolved stellar populations in an NSC in great detail.  Such an analysis is not feasible for the Milky Way's galactic center or other nearby NSCs like M54  because of their large distances as well as high levels of crowding and extinction, which make it observationally very expensive to obtain high-resolution spectra for a large number of stars in a comparable way to \ocen.  More generally, \ocen\ is a prototype of multi-population globular clusters, a class of important astrophysical systems whose evolution remains to be understood.

Chemical elements created inside stars, or during stellar explosions, and ejected into the interstellar medium on varying timescales and in different amounts \citep{Kobayashi20, Cowan21}. These ejected elements can then become locked back into new stars, formed from this chemically-enriched interstellar gas. The chemical abundances of low-mass (long-lived) stars therefore serve as a fossil record of their birth environments, and, as such, are essential for studying galactic evolution \citep{Freeman02, Tissera12}. Critically, chemical abundances in stars trace all the key nucleosynthetic element-creation paths, namely $\alpha$-capture, iron-peak, and various neutron-capture modes. 
Thus, the more chemical elements we have available for analysis in each star in these fossil records, the better we can disentangle the evolutionary timescales and processes \citep{Griffith24}.  Today, there are many different chemical abundance measurements of thousands of individual stars in \ocen\ published \citep{All_ctio, All_apo}. From these datasets, many studies have attempted to use clustering methods to find groups in these stars and chemical elements to interpret the main characteristics \citep{All_apo, AlvarezGaray24}, yet no definitive conclusions regarding the history of \ocen's populations have been reached. One reason might be because these studies are rather focused to separate the stars into distinct groups, and not necessarily designed to study how these groups connect in an evolutionary and historical sense and are thus related to each other in an astrophysically viable evolutionary sequence.

Biology has an extensive track record developing and implementing phylogenetic tools to study relationships among species and so reconstruct their histories. Phylogenetic trees study the accumulation of small modifications passed from one generation to the next to build a model of their relationships as they developed over time  \citep{baum2005tree}. The key assumption in the use of trees is that there is a level of heritability between generations. In biology that heritability is provided by DNA. Indeed, it is the fact that DNA is shared across all organisms that allows the reconstruction of relationships across all of life, and the expansion of genomics techniques has revolutionized phylogenetic methods \citep{yang2014molecular}. DNA carries information about shared ancestry, and large genomic datasets have been used to model and analyze trees in evolutionary biology.

However, although DNA is the basis for most phylogenetic reconstructions in biology, as long as there is a mechanism of heritability then other systems can be used \citep{o2025archaeology}. All that is required is some means by which information (and traits) are transmitted from one generation to another. For example, languages are inherited (and subject to modification from generation to generation), and have been analyzed  phylogenetically \citep{bromham2022global}. Archaeologists increasingly use these techniques to reconstruct cultural history from non-biological materials \citep{o2025archaeology}. Phylogenetic methods have also been applied to nearby stars using chemical abundances as heritable traits \citep{JofreTree}. Indeed, stars generate new chemical elements themselves, which are expelled into the interstellar medium upon their death. New stars form from this chemically enriched interstellar medium, inheriting its chemical pattern. This fullfills the heritability condition required for phylogenetic methods. Some low-mass stars survive for as long as the age of the Universe, and their chemical pattern serves as the fossil record and heritable marker of the evolving interstellar medium \citep{Freeman02}. It is thus possible to build on approaches from evolutionary biology, adapt tools to reconstruct the history of galaxies by using the chemical patterns of stars as the heritable marker, and illustrate the power and opportunity of such inter-disciplinary approach \citep{JofreTree, Jackson21, DaniTree}. The complex nature of \ocen, and the wealth of available stellar abundance data, make it an ideal case for such research.

In Sect.~\ref{sect:data} we present the data used in this work, which we take from published catalogs. In Sect.~\ref{sect:method} we explain the phylogenetic methods used to analyse \ocen, and in Sect.~\ref{sect:ocen} we provide an astrophysical interpretation of these results. We conclude our work in Sect.~\ref{sect:concl}.

\section{Data}
\label{sect:data}

In this work we use two published datasets of stars from \ocen, both containing data of chemical abundances obtained from high resolution spectral analysis. The idea to use both datasets here is to further investigate the impact of the choice of data on our conclusion. These datasets come from \cite[][hereafter Optical]{All_ctio} and \cite[][hereafter Infrared]{All_apo} and we explain them in more detail below.  Additionally, we checked that both the Optical and Infrared datasets with detailed abundances are highly probable members of \ocen, according to \cite{VasilievBaumgardt21}. This is based on the updated Gaia EDR3 kinematic properties of Milky Way clusters. Both datasets (IDs, brightness, and abundances) are listed in the CDS tables.

\begin{figure*}[t]
\centering
\includegraphics[scale=0.5]{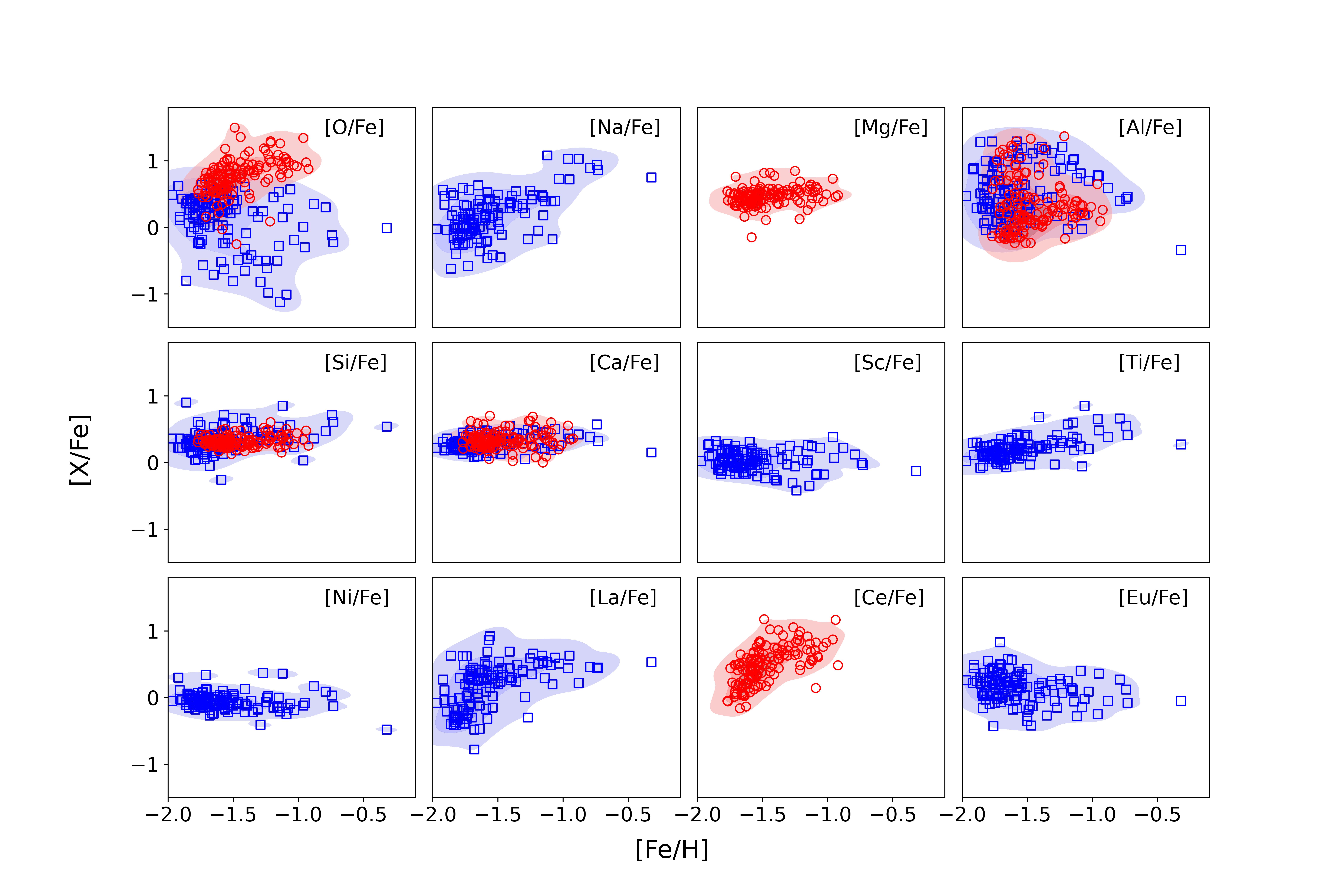}
\caption{Abundance ratios against [Fe/H] of all elements in both optical (in blue squares) and IR (in red circles) samples. Some of the element abundances are measured in both samples, and some are measured in only one of the samples.  Shaded contours correspond to the Gaussian kernel density estimation of the datapoints.}
\label{fig:abund}
\end{figure*}

\subsection{Optical sample}

\cite{All_ctio} published stellar parameters and abundances for 855 red giants determined from spectra taken from Cerro Tololo Inter-American Observatory (CTIO) using the Blanco 4~m telescope equipped with the Hydra multifiber positioner and bench spectrograph. The spectra cover the wavelength range of $6135 - 6365$ \AA\  and $\sim 6500 - 6800$ \AA, and have a resolving power of $R \sim 18,000$. The long exposure of the observations yielded high signal-to-noise spectra of about 200. The authors performed a spectroscopic analysis based on equivalent width methods to determine abundances of Fe, Na, Si, Ca, Sc, Ti, Ni, and Eu. Abundances of O, Al, and La were determined using synthesis fitting. For Sc abundances of both ScI and ScII were provided, which we take the mean of.  Among these 855 stars, 122 stars have abundances measured for all the elements. We consider this subsample for our analysis. 

Photometry, coordinates, and membership probabilities for
all stars in that study were taken from \cite{vanLeeuwen2000}. This is before the Gaia era, thus the proper motions were derived using photographic plates from the years 1931, 1937, 1978, and 1983. Because of this, obtaining updated photometry and astrometric information from Gaia for these stars was possible for only 710 stars, including all our 122 selected stars.  
The stars are in the red giant branch, except a few outliers \citep[see the study of multiple red giant branches in \ocen\ by][]{Pancino2000}.

\subsection{IR sample}

This sample is taken from \cite{All_apo}, who analyzed stars observed with SDSS-IV/APOGEE-2 \citep{apogee2}. It consists of 775 stars with APOGEE observations, which use an infrared spectrograph in the H band with a resolving power of $R \sim 22,000$ covering a range of 1.514-1.696 micrometers installed in the du Pont Telescope at Las Campanas Observatory. The data for \ocen\ has a signal-to-noise of about 70.  Using the code BACCHUS \citep{bacchus}, \cite{All_apo} published abundances of Fe, C, N, O, Mg, Al, Si, K, Ca, and Ce. The subset which has measurements for all elements contains 137 stars and this is the set we use for our analysis. In this work we remove C and N abundances because in red giants they change as stars experience their dredge up \citep{Salaris15, All_apo}, making the photospheric abundances of these stars not a suitable heritable tracer.

Cluster membership for these stars is based on RV and proper motions from Gaia DR2 \citep{Baumgardt18}.  
The cross-match of this sample with Gaia DR3 yields 676 stars, with the 137 stars with detailed abundances included. 
As the Optical sample,  they follow the red giant branch. There are a few stars in the Infrared sample which are hotter and of lower luminosity compared to the Optical sample, but both datasets are otherwise comparable.

\subsection{Chemical abundances}

Figure~\ref{fig:abund} displays the chemical abundance ratios for all elements analyzed in this work in various panels. The panels show the distribution of each abundance ratio [X/Fe] as a function of [Fe/H].  In blue squares we plot the abundances for the Optical sample, and in red circles we plot the abundances for the Infrared sample. A total of 12 abundance ratios, in addition to the metallicity [Fe/H] are included in this analysis. 

With the exception of a few stars from the Optical reaching values up to $\mathrm{[Fe/H]} = -0.3$ dex, both samples cover a comparable metallicity range between $-2$ and $-1$ dex. We only have abundances of O, Al, Si, and Ca for both samples. All these elements are mainly produced in core-collapse Type II supernova \citep{2013Nomoto}, although Al is very dependent on the progenitor metallicity \citep{2019Weinberg, vasini24} and has a different behaviour compared to the $\alpha-$capture elements like O, Si, and Ca.  Mg, which is another $\alpha-$capture element is only included in the IR sample, while Ti, which shows the trends of an $\alpha-$capture element in the Milky Way, is only included as part of the optical sample.  Ca, Mg, and Si, the other $\alpha-$capture elements of the sample, show a flatter trend as a function of metallicity. We remark that among these elements, O, Na, Mg, Al, and Si are those that show high levels of depletion or enhancement due to the internal stellar evolution processes of the CNO, NeNA, MgAl cycles that are suspected to produce the anticorrelations in GCs \citep{gratton19}. 

Al, Si, and Ca agree well in both samples, showing similar trends and absolute values. This is not the case for O, which shows a positive trend with [Fe/H] and very high abundances in the Infrared sample, while the [O/Fe] trend with [Fe/H] decreases with [Fe/H] in the Optical sample. This discrepancy is discussed further the next section.  
The high [O/Fe] abundances of some stars from the Infrared sample and its increasing trend with metallicity 
have also seen in the independent analysis of \cite{Marino12}, who analyzed  a sample of \ocen\ stars observed with optical spectra.  We recall that it is not expected that the plausible NSC nature of \ocen\ has similar abundance patterns as the Milky Way or disrupted dwarf galaxies.
Indeed, \cite{Romano07} predict high [O/Fe] abundance ratios in \ocen, although with a decreasing trend with metallicity. Star formation bursts can cause an increase for [O/Fe] in dense systems, this was recently shown in the EDGE simulation of a NSC by \cite{Gray24}. 
We keep thus all [O/Fe] for our analysis. 

The Optical sample includes, in addition to Fe, the iron-peak abundance ratios Sc and Ni. Both Sc and Ni show flat trends and a small dispersion with [Fe/H], although Sc displays a small decrease over metallicity as it is being mostly produced by core-collapse supernovae \citep{Kobayashi20}.  Other elements produced in massive stars, which are only included in the Optical sample are Na and Eu. Both show trends with [Fe/H] and a high dispersion. Europium however is also produced by neutron star mergers \citep{Cowan21, arcones23}, which are able to enrich entire dwarf galaxies with high levels of Eu in one episode of star formation.  Finally, both samples have one neutron$-$capture element produced by the $s-$process and expelled to the interstellar medium via asymptotic giant branch (AGB) winds \citep[see for instance][]{Lugaro03, Lugaro12, Cseh18,2018Magrini}. The Optical sample includes La, and the Infrared sample includes Ce. Both elements show comparable trends, such as increasing with [Fe/H] from about -0.5 to 1 dex. This is expected  \citep{2014Bisterzo} because such elements produced in AGB stars are typically released after SNIa elements. Their trends indicates an extended star formation history.

\subsection{Comparison of common stars between samples}

\begin{figure*}[t]
\includegraphics[scale=0.6]{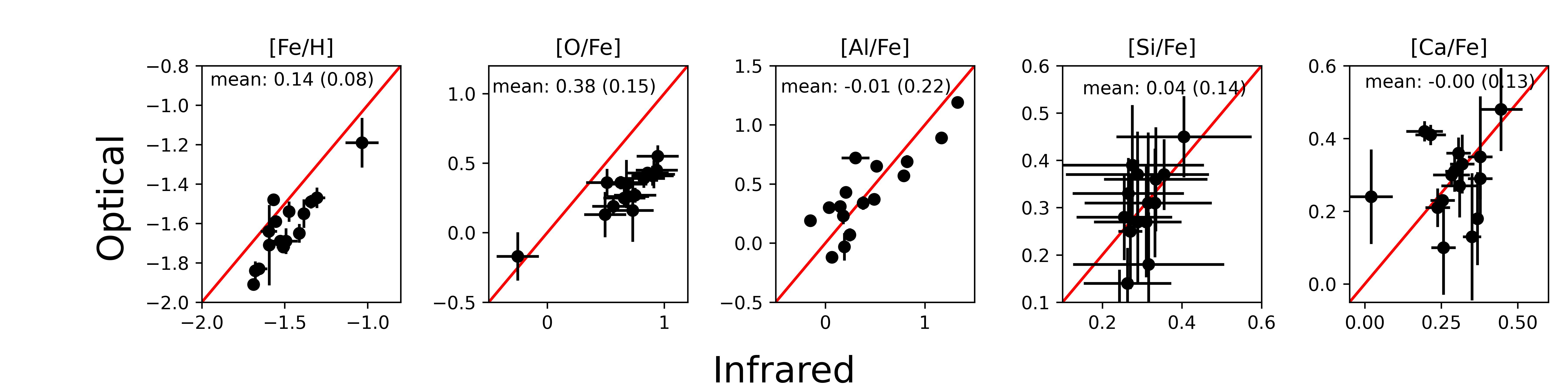}
\caption{Comparison of abundance ratios for the 16 stars in common between both datasets. The mean and the standard deviation of the comparison is indicated in each panel. The diagonal red line indicates the 1-1 relation. Only [Fe/H] and [O/Fe] show significant systematic offsets. }
\label{fig:chemical_comp}
\end{figure*}

Our cross-match with Gaia allows us to identify stars in common between both samples, finding 16 stars. Their IDs and Gaia magnitudes are listed in Table~\ref{tab:sipsop}. In addition to the Source ID from Gaia, we include the APOGEE ID used by \cite{All_apo}, the LEID ID used in \cite{All_ctio}, and our own IDs, which we call SIR for the Infrared sample, and SOP for the Optical sample.

The direct comparison for the abundances in common can be found in Fig.~\ref{fig:chemical_comp}. The x axis show the abundances as reported by \cite{All_apo} and the y axis show the abundances of the same stars as reported by \cite{All_ctio} for [Fe/H], [O/Fe], [Al/Fe], [Si/Fe], and [Ca/Fe]. We overplot the one-to-one line in red and write in each panel the mean of the differences and the standard deviation of the differences. We note the small errors reported for Al. Indeed, for most stars, no Al errors are reported in \cite{All_ctio}. This is because in that work errors are the scatter among lines. Measuring Al from optical spectra is very challenging (see for instance \cite{Buder+2022}), the Al abundances here come mostly form one line, which is why no errors were reported.  
While the agreement for Al, Si, and Ca are within the scatter, there is a systematic offset of 0.15 dex in metallicity between both samples, and a significant offset in [O/Fe] of 0.4 dex. 
The large difference in [O/Fe] between optical and infrared measurements in metal-poor stars has been already identified in the literature \citep{Griffith19}. While applying a shift might bring into better agreement both samples in the [Fe/H] - [O/Fe] plane, the trends still would go on opposite directions.  It is thus not possible to fully attribute the differences in oxygen between samples to a systematic effect; the difference might also be due to a selection effect of stars. 

\begin{figure*}
\centering
\includegraphics[scale=0.6]{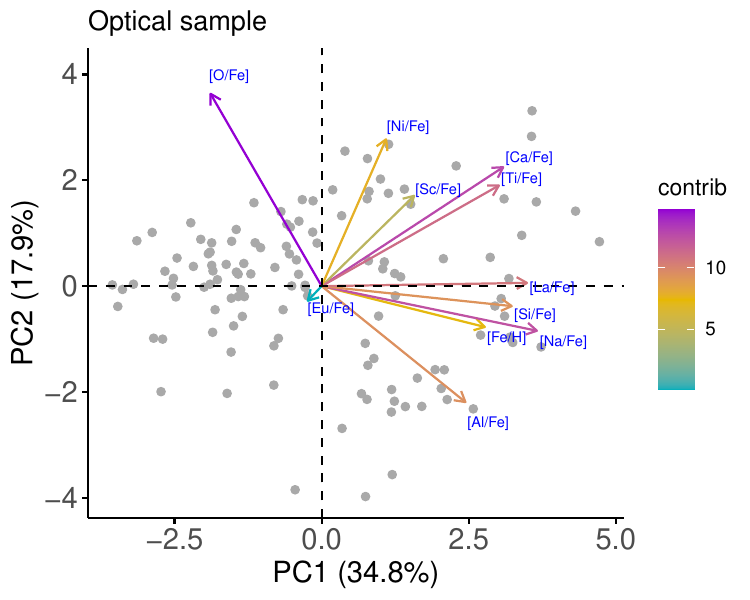}
\includegraphics[scale=0.6]{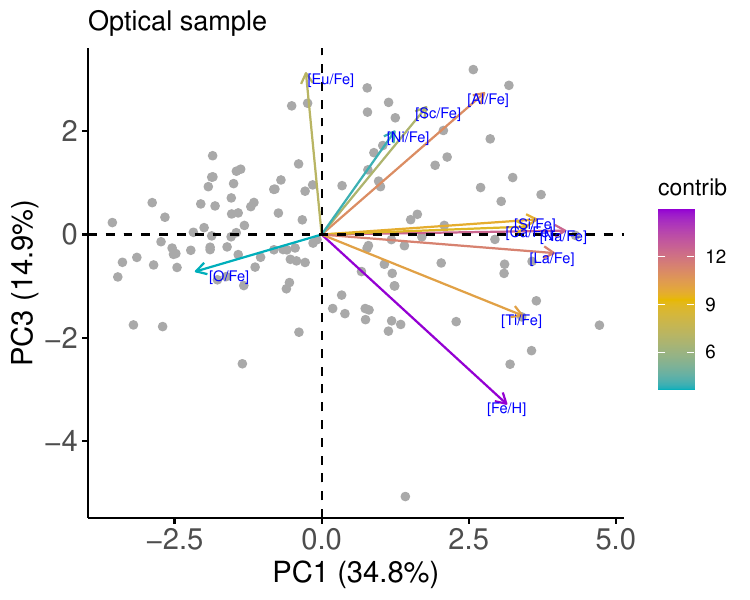}
\caption{PC space with the direction and contribution of each chemical element in the first three principal components.  Stars are dots in the space and arrows indicate the direction in which the abundance ratios are higher in these planes.  The color of an arrow shows the overall contribution of the chemical element to both PCs. The first two components show how $\alpha$-elements and [Al/Fe] dominate the variance. These arrows point in opposite directions in the space. This is due to the anticorrelation of these elements in the chemical space. The third component illustrates that [Fe/H] and [Eu/Fe] become important. }
\label{fig:pca_planes}
\end{figure*}

\section{Phylogenetic analysis}\label{sect:method}
Our goal is to use chemical abundances to study the assembly history of stars in \ocen. This is analogous to biologists studying the genealogical history of a biological population, such as asking if a human population is descended from one or multiple sources through migrations and admixture\footnote{ In population genetics, migration refers to the movement of genes from one population to another. This can occur when individuals migrate and introduce their genetic material into a new population. Admixture refers to the mixture of previously isolated populations. This can happen between different species (hybridization) or within the same species.} \citep{hellenthal2014genetic}. This approach that biologists commonly use can be 
adapted to study the ancestry of stars. This process usually starts with a principal component analysis (PCA) to visualize the variability in the data, followed by a mixture model to detect distinct groups in the data, and finalized by a phylogenetic analysis to reconstruct when and where migration and admixture events happened \citep{nespolo2020out, price2006principal, Tian08, solovieff2010clustering}. In this study, we adapt this approach to help us understand the formation of \ocen. We focus our analysis and discussion on the results obtained from the Optical dataset, but the parallel results for the Infrared sample can be found in the appendixes.

First, we apply PCA to our dataset. This is to ensure that there is chemical diversity in the dataset to detect distinct groups \cite{Buckley24} and to understand which elements are responsible for this diversity. PCA has been done with chemical data of Milky Way stars \citep[][]{Ting12, Andrews12PCA, Signor14, Buckley24, Griffith24}.  In Fig.~\ref{fig:pca_planes} we show the distribution of 122 \ocen\ stars in the space defined by the first three principle components (PCs), and the contribution of significant chemical abundances to each PC. The left-hand panel shows the first and second component, while the right-hand panel shows the first and the third component. A similar plot but for the Infrared sample can be found in Fig.~\ref{fig:pca_planes2}. Each star is represented by a gray circle.  The arrows of each abundance ratio are overplotted in the diagrams whose length and color correspond to the contribution in these dimensions while the direction of the vector corresponds to where the stars have a higher abundance ratio in each PC.  These planes in a way summarize Fig.~5 of \cite{Buckley24}, who plotted the first and second component of the PCA color coded by different abundance ratios in nine panels. 

The plane shows that an anticorrelation of aluminum and oxygen, as well as sodium and oxygen, have an important impact on the variance of the data. This can be seen through the [Na/Fe] and [Al/Fe] vectors pointing in the opposite direction to the [O/Fe] vectors.  Indeed, the Na-O and Al-O anticorrelations are typical of GCs \citep{Carretta09}, whose nature remains unresolved \citep{BastianLardo18, gratton19},  and have been studied in \ocen\ \citep{All_ctio, AlvarezGaray24}.  We see long arrows for those elements pointing to opposite directions. The first PC captures 34.8\% of the variability among stars in their chemical composition, which is mostly driven by the Na-O anticorrelation, with some influence from  La, Si, Ca, Ti, and Fe (see Sect.~\ref{sect:pca}). The second PC (Y axis of left hand panel), captures 17.9\% of the variance and separates the stars mostly by their $\alpha$-capture (O, Ca), iron-peak (Ni) elements and aluminum. The third dimension (Y axis of right-hand panel) contributes 14.9\% to the variance and separates the stars in neutron-capture (Eu) abundance ratios as well as iron. For more in-depth analysis of the PCA, as well as the PCA using the Infrared sample, we refer the reader to the Appendix~\ref{sect:pca}. When using the Infrared sample (see Fig.~\ref{fig:pca_planes2}, the [O/Fe] and [Al/Fe] vectors point in the opposite directions to the Optical sample (reflecting the different trends with [Fe/H]), but the  O-Al anticorrelation still holds. 
We conclude that we need to consider all available elements because they all contribute meaningfully to at least one of the three most significant principal components.

\subsection{Stellar population bar plot}
Given that we have found that our data is chemically diverse, we proceed to cluster the stars using all abundance ratios, [X/Fe], with Gaussian mixture models (GMM).
The Bayesian Information Criterion (BIC) scores indicate that the optimal number of units or groups in this dataset is three (see Appendix.~\ref{sect:gmm}). In Fig.~\ref{fig:structure_gmm} we show the GMM classification for three groups for all stars. 
This plot is similar to the {\tt STRUCTURE} bar plot, which is a popular tool for studying population structure in biology \citep{structure}. While {\tt STRUCTURE} uses a mixture model for DNA, which is a categorical variable, we adapt the idea to describe stellar populations by using GMM for chemical abundance ratios. We call our approach {\it stellar population bar plot}.

\begin{figure*}[t]
\vspace{-1cm}
\centering
\includegraphics[scale=0.29]{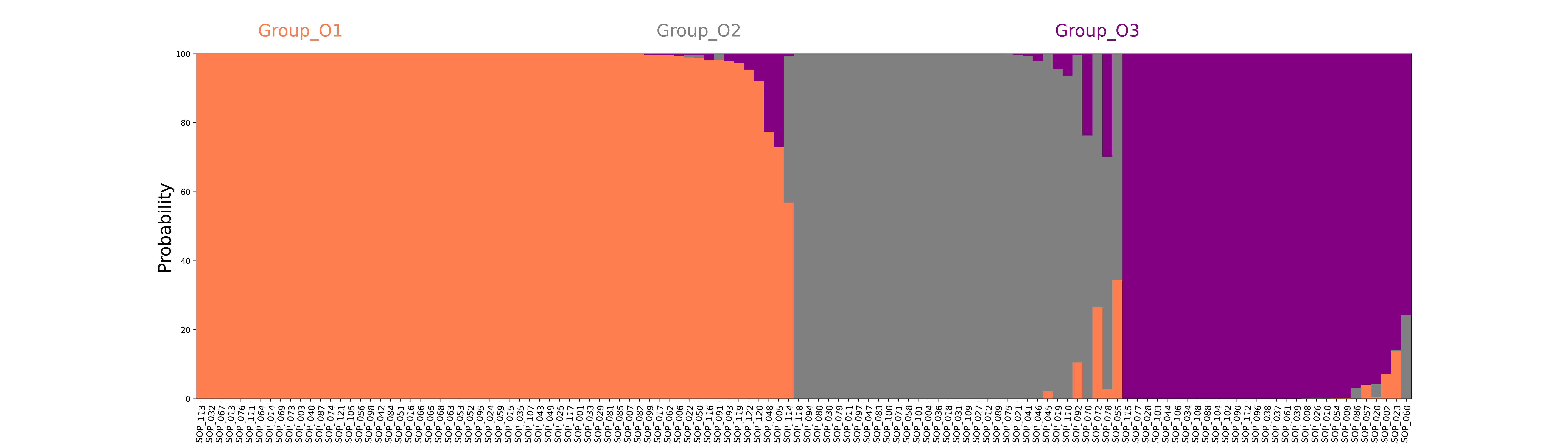}
\caption{Stellar population bar plot. Each star, whose name is indicated in the X axis, is shown as a bar with the cumulative probability of the star belonging to a specific group. The Y axis shows the probability of the star belonging to the groups, which are assigned using a Gaussian mixture model of three components.  The stellar population bar plot allows us to appreciate the clustering, and to identify which groups might be more related to each other, as reflected by the level of mixing between them. }
\label{fig:structure_gmm}
\end{figure*}

The main difference between the stellar population bar plot and how astrophysicists usually use GMM \citep{Das+2020, Buckley24, Buder+2022} is that we do not only select stars with high probability of belonging to a certain GMM group for further analyses. Instead, the bar plot includes stars with all levels of probabilities. This is because stars with a low probability of belonging to a certain group offers us the opportunity to study mixing between groups \citep{villarreal2022identification, nespolo2020out}. This is similar to the {\tt STRUCTURE} bar plot where low probability can be indicative of migration or admixture events, for example when a particular population is descended from multiple ancestral populations  \citep{peter2016admixture}. If the majority of stars have a 100\% probability of belonging to one GMM group, then the distribution of chemical abundances over stars is discrete, suggesting that each GMM group is likely to have evolved independently. But if the majority of stars have a low probability of belonging to a single group, then the distribution of chemical abundances over stars is continuous. This suggests strong mixing among groups, for example, the chemical elements from one group is mixed with the interstellar medium that forms the other group.

Most of the stars with a high probability of being  in Group\_O1 (orange) have very low probabilities of being in another group. This suggests that Group\_O1 has evolved independently of the other groups.  This independency might be spatially or temporally, namely this group could have formed far away from the other groups and/or before the other groups existed. On the other hand, the gray group has a handful of stars with more than 20\% of contribution from the orange and purple groups, suggesting that the gray group could be more related to the others.  This interpretation is in any case subject to the uncertainties of the abundance measurements, as well as the latent chemical space considered for the clustering \citep{Buder+2022, Buckley24}, and also the number of GMM groups. To draw firm conclusions, we would need validate with simulated data, such as \cite{Gray24}, where the history is known.  Various key abundance ratio planes color-coded by the GMM groups can be found in Appendix.~\ref{sect:chemical_distr}, where they are discussed at length.

\subsection{Phylogenetic tree of \ocen}\label{sect:trees}

While phylogenetic trees can show when and where migration or admixture events happened in the history of a biological population, they also carry important information on the history of star formation. In particular, using simulations of galaxy evolution \cite{DaniTree} showed that an isolated galaxy generates stars that form a special tree topology, known as the `caterpillar' tree, where all tips are incident to the same branch. We can therefore use this result as the benchmark to compare the tree of \ocen\ stars with the null hypothesis that \ocen\ evolved in isolation.

\begin{figure*}[t]
\centering
\includegraphics[scale=0.4]{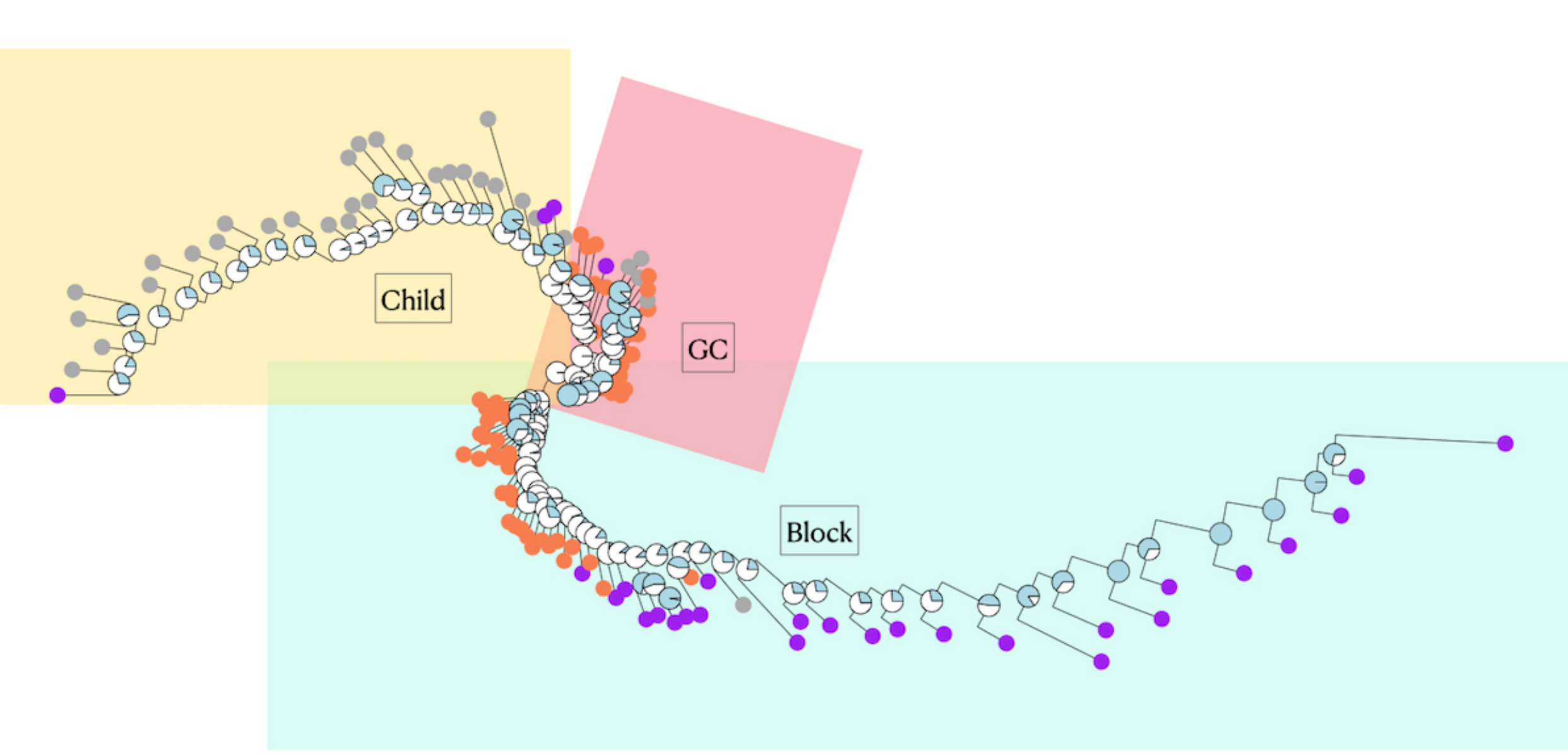}
\caption{Circular Optical tree. The colors of the tips correspond to the GMM groups. The pie charts on the nodes correspond to the percentage of occurrence of the node in 1000 random generated trees considering the abundance uncertainties (node support), indicating that the tree has an overall support of 35\%.}
\label{fig:optical_unrooted_tree}
\end{figure*}

\subsubsection{Building NJ trees}
Following \cite{DaniTree}, we built trees using the empirical distance method and the agglomerative neighbor-joining (NJ) clustering algorithm \citep{Saitou1987}, which does not rely on an evolutionary model to find the best tree from a given dataset. This is important when borrowing tools from a different discipline, as we must avoid making assumptions that might not apply to astrophysics.  The NJ algorithm arranges chemically similar stars nearer to one another in the tree while placing those more different further apart. The branch lengths correspond to the overall  chemical  differences between stars \citep[see also Sect. 3 of][]{JofreTree}. The chemical distance is taken from a distance matrix, which is symmetric and of the dimension of the number of stars. In this case, following \cite{JofreTree}, we computed Manhattan distances based on stellar abundances. The pairwise distance matrix was calculated using the [X/H] abundance ratios available to us, namely [O/H], [Na/H], [Al/H], [Si/H], [Ca/H], [Sc/H], [Ti/H], [Fe/H], [Ni/H], [La/H], and [Eu/H].

To assess the robustness of our tree topology we calculate the node support by considering the uncertainties in our abundances. We sampled 1000 distance matrices, by  perturbing the abundances with their uncertainties. From these distance matrices we calculated 1000 NJ trees.  The node support corresponds to the percentage of occurrence of a given node in all other sampled trees. This is calculated by counting the occurrence of nodes in all trees \citep[see][for more details about node support in stellar trees]{Jackson21, walsen24}. 

\subsubsection{Stellar phylogenies}
Figure~\ref{fig:optical_unrooted_tree} shows the tree in circular form and allows us to realize that the tree shows similar results to the GMM analysis in the sense that purple and gray stars form two `caterpillar' trees, suggesting they evolved independently.  Since these branches contain stars from the different GMM groups, we have labeled with new names, which are explained in detail in the next Section.  The branches are highlighted in the yellow and cyan boxes.  The remaining stars, highlighed in the pink box, form a more `star-like' tree, suggesting that they either mixed a lot,  are from different sources,  or the data is too uncertain for the NJ method to resolve the hierarchical differences there. The nodes of the figure have pie charts overplotted, which indicate the proportion of trees that have this node in blue given the uncertainties in the abundance measurements (this is the node support). Due to uncertainties in abundance measurements, multiple pair-wise distances between stars in their chemical composition are possible, which translates in an uncertain tree topology.   It is possible to see that the orange stars, which belong to the Group\_O1 from the GMM and are mostly part of the pink and the deep nodes of the Block box, have nodes with low support. This is because these stars are so similar to each other that the NJ algorithm is not able to assign them to a specific node given the uncertainties. Indeed, here we find ourselves with a limitation of the NJ method, already discussed in \cite{walsen24}. By construction, the NJ has to assign every star into a different branch, assuming each of them represents a different star formation episode. When analyzing observed stars, we cannot rule out the possibility that two stars belong to the same star formation episode. While the GMM will put them together in the same group (because they might have the same chemical signature), the NJ algorithm has to place them in distinct leaves. The overall support of 35\% is similar to \cite{walsen24}, which is expected since this is the support due to uncertainties in the stellar chemical data. 

\section{The assembly of populations in \ocen}\label{sect:ocen}

\begin{figure*}[h!]
\centering
\includegraphics[scale=0.18]{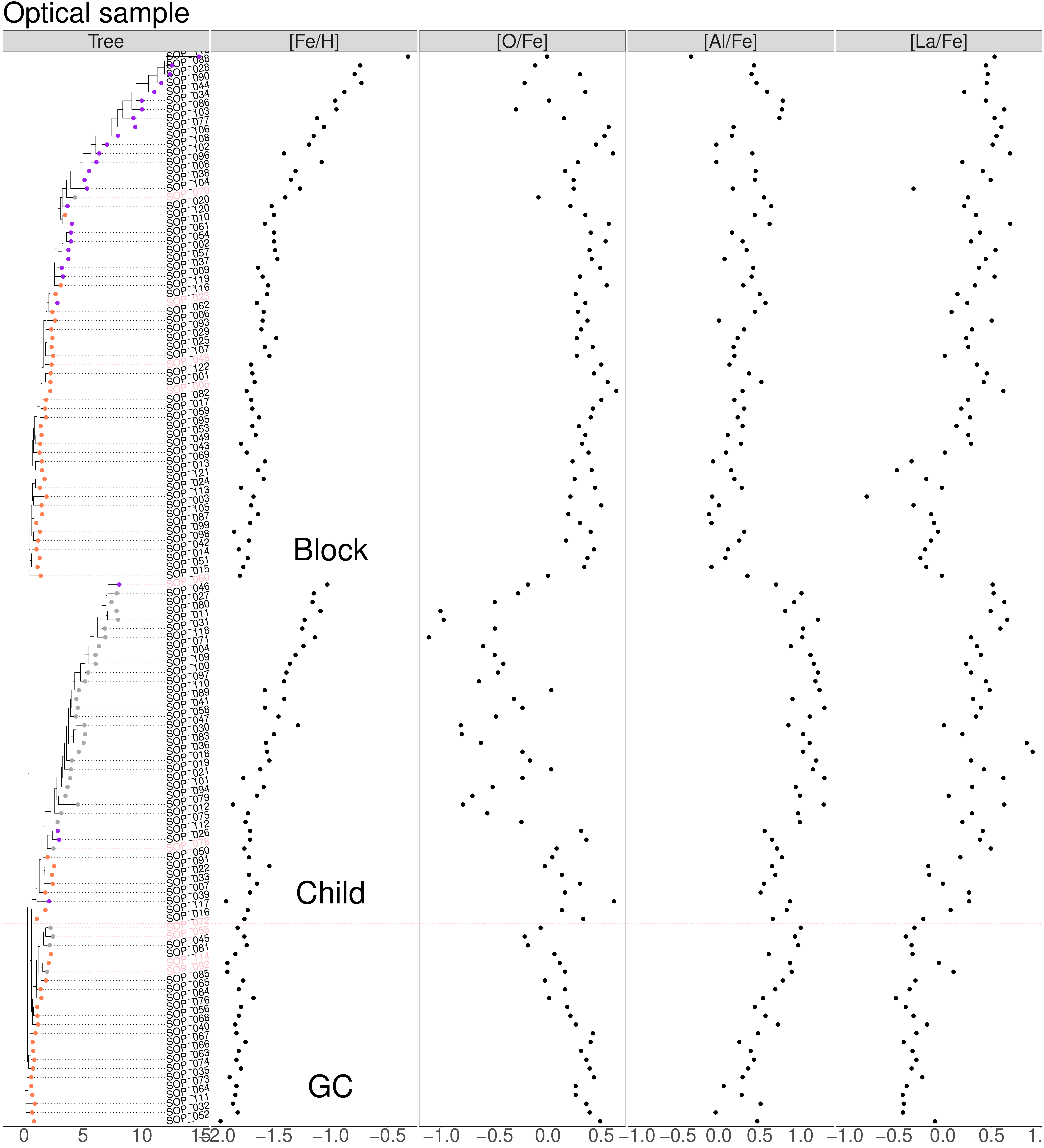}
\caption{Neighbor-joining tree of \ocen\ stars. Tips are colored based on the GMM group to which each star has the highest probability of belonging to in Fig.~\ref{fig:structure_gmm}. The tree shows three main branches, whose [Fe/H], [O/Fe], [Al/Fe], and [La/Fe] are shown in the right-hand panels. The GC branch has properties similar to globular clusters. The Block and Child branches have extended metallicity and [La/Fe] distributions, supporting in-situ star formation for a prolonged duration. However, their [O/Fe] and [Al/Fe] trends are distinct, suggesting that they had different star formation histories and explaining why they are placed on different branches and GMM groups.}
\label{fig:tree}
\end{figure*}

When the root, that is,  the ancestral node, is set, and assuming the choice of root is sensible \citep{yang2014molecular}, the phylogenetic tree reveals more information on star formation history. The resulting rooted tree is displayed in Fig.~\ref{fig:tree}. We set the root at the star with the lowest [Ca/H] abundance ratio. This assumes that chemical abundances increases with time, so the lowest abundances in a dataset should reflect the earliest time and therefore a root for the tree. We choose [Ca/H] instead of [Fe/H] in this case because of its excellent consistency in the abundance measurements with the Infrared sample, which allows us to use the same criterion to root the Infrared tree.  Furthermore, the PCA revealed that Ca had a similar effect on the chemical diversity of both samples. This enables a better comparison between trees and so a better interpretation of our results (see Sect.~\ref{sect:trees}).

 Our analysis revealed that stars with pink labels stars, that is, stars whose probability to belong to a specific group is lower than 90\%, do not have a preferred location in the tree. Changing the threshold for this definition does not affect our interpretation.  In this section we interpret the astrophysical meaning of the branches and groups found in our phylogenetic analysis, and so discuss the possible history of assembly of \ocen.

\subsection{The Block, Child and GC branches of the tree}. Figure~\ref{fig:optical_unrooted_tree} and \ref{fig:tree} show two big branches, each with a clear `caterpillar' topology. We call these two branches {\it Block} and {\it Child}. The remaining tips are closer to the root of the tree and show a more star-like topology, which we call {\it GC}. Considering our null hypothesis \citep{DaniTree},  the Block and Child branches likely evolved independently, while group GC likely consists of stars well mixed or from different origins. The chemical patterns in Block, Child, and GC groups are not only distinct, but also suggest their possible history. The next four panels to the right of Fig.~\ref{fig:tree}  show the distribution of [Fe/H], [O/Fe], [Al/Fe], and [La/Fe] along the branches.  Evidence supporting Block and Child having independent star formation histories is that the two branches have  different [O/Fe], [Al/Fe], and [La/Fe] trends. Also, the  majority of purple stars, grouped by GMM, are in the Block branch and the majority of gray stars are in the Child  branch. In contrast, GC stars have a narrow [Fe/H] and [La/Fe] ranges, and present the Al-O, Na-O anticorrelation, which is typical in globular clusters, and so we call this population {\it GC-Cen}. 

The population of stars from the Block branch shows a classical [O/Fe] trend with metallicity, namely the [O/Fe]-[Fe/H] relation  with a [O/Fe] plateau at 0.4 dex, which is the signature of SNII and presenting a potential [O/Fe] knee at $\mathrm{[Fe/H]} \sim -1.3$,  which results from the pollution of [Fe/H] due to SNIa. This is consistent with the chemical evolution of a dwarf galaxy \citep{matteucci90}. The increasing trend of [La/Fe] with [Fe/H] is also found in dwarf galaxies.  Since La is a neutron-capture element produced via the slow-process mechanism in AGB stars, reaching the high [La/Fe] abundance ratios is possible only after a few Gyr \citep{2014Bisterzo, Romano23}.  
[Al/Fe] is notably more enhanced than in typical dwarf galaxies, which are normally subsolar and lower compared to the Milky Way trend \citep{Das+2020, 2021Hasselquist}. We note however that recent discoveries of Milky Way metal-poor ancient structures, such as Shiva and Shakti \citep{MalhanRix24}, as well as the G3/G5 groups from the GALAH sample by \cite{Buckley24}, have significantly higher [Al/Fe] than the better-known disrupted galaxies in the halo of comparable metallicities. Such systems were proposed by \cite{MalhanRix24} to have formed from gas clumps in a massive and/or dense progenitor. The abundance patterns of the Block branch could be thus associated with a building block population or a proto-galaxy, equivalent to Shiva and Shakti of the Milky Way. We thus call this population {\it Block-Cen} and attribute it to the building block of the progenitor galaxy that formed \ocen.  

The Child branch has an anticorrelation between Al and O like the GC branch, although [Fe/H] has a significant increasing trend, as does [La/Fe]. This points towards an extended star formation history, perhaps a result of in-situ star formation from material that has been enriched by the stars within \ocen, henceforth its designation as {\it Child-Cen}. Indeed, the G\_O2-gray stars has slightly more admixture with G\_O1-orange and G\_O3-purple in the stellar population bar plot, suggesting that this group is somehow more related to the other populations. This admixture is driven by [Ca/Fe] and [La/Fe] abundances (see Fig.~\ref{fig:chemtrends_gmm}). 

 The anticorrelations of stellar abundances such as [O/Fe] and [Al/Fe] are among the most challenging features of globular clusters to be explained \citep[see][for extensive review]{gratton19}. While there are strong arguments in favor of an inheritance process due to the CNO and MgAl cycles inside massive stars, which pollute the ISM in the AGB phase via winds \citep[][]{dercole12, gratton19}, there are still many theoretical aspects on stellar nucleosynthesis and GC formation that are not understood and therefore the observations of such abundance ratios cannot be fully reproduced for all GCs \citep{BastianLardo18}. The case of \ocen\ offers additional complexities due to the extended range of [Fe/H] and s-process elements, which suggest timescales of Gyr for star formation compared to the few Myr of the CNO, NeNA, and MgAl cycles of massive stars. The peculiar abundance patterns of {\it Child-Cen}, namely the very low [O/Fe], and high [Al/Fe], cannot be directly associated with the (unexplained) O-Al anticorrelation of classical GCs, because of the different timescales needed to simultaneously describe the pollution mechanisms of the anticorrelations and the chemical enrichment of the Fe and La trends. A stellar population with such characteristics has not been found in the Galactic field, or in mono-metallic clusters. The chemical pattern of {\it Child-Cen} would therefore need a new explanation.

\subsection{Assessing systematic differences with common stars}\label{sect:trees}

\begin{figure*}[h!]
\centering
\includegraphics[width=\linewidth]{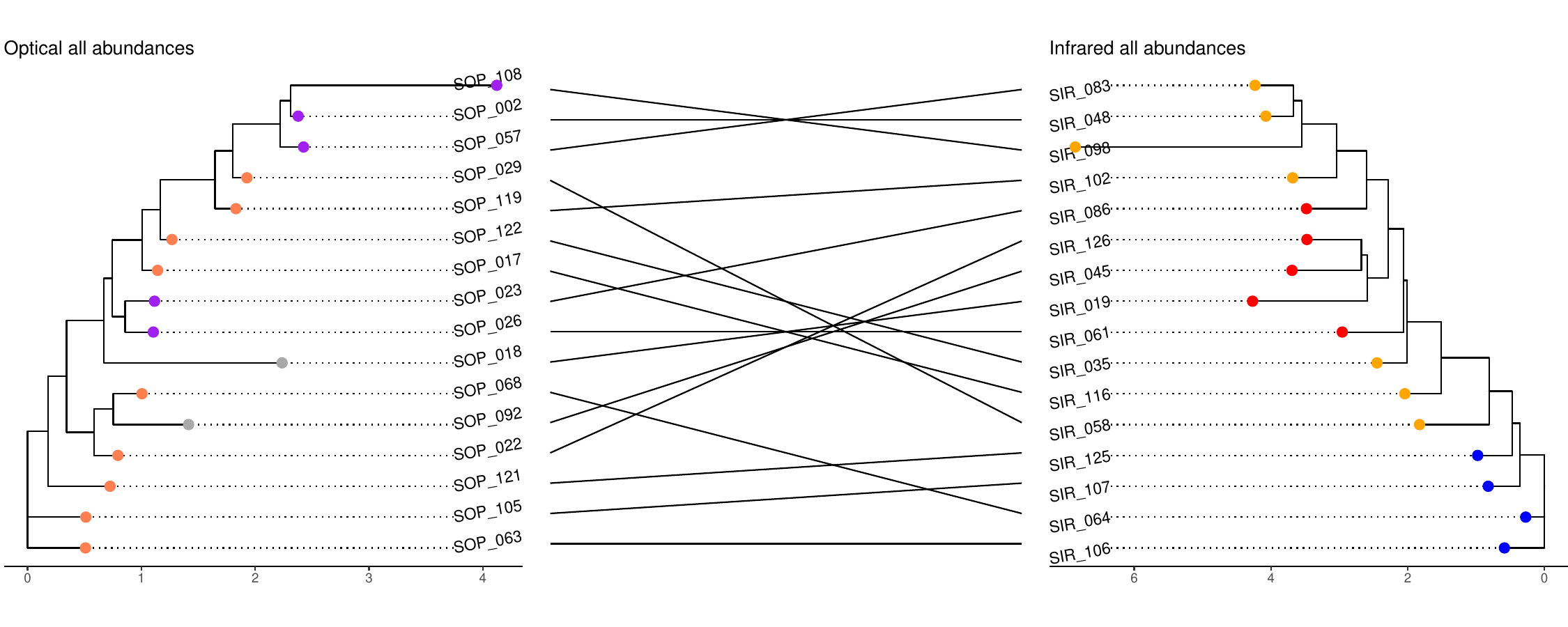}
\includegraphics[width=\linewidth]{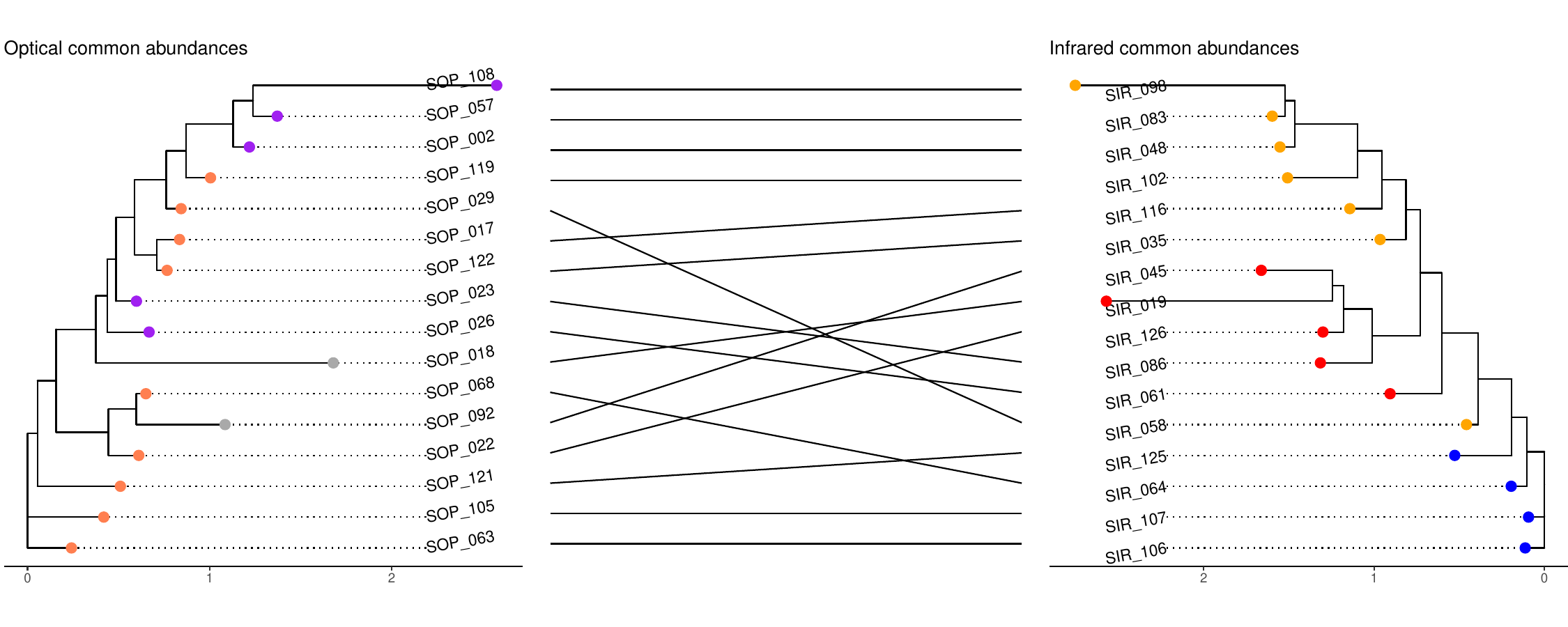}
\caption{Comparison of rooted tree color coded by groups for common stars. Top Panel shows the comparison among trees when all abundances are considered for building the tree, while lower panel shows the comparison when elements in common are used.  The agreement between trees is slightly better when the same chemical abundances are used, but differences persist. This can be attributed to the uncertainties in the data.  }
\label{fig:both_trees}
\end{figure*}

To understand further the similarities and the differences in the populations and branches found between both datasets (see Appendix.~\ref{sect:ir_tree} for the tree obtained from the Infrared sample), as well as the tree shapes, we build trees of the stars in common between both datasets.  The trees are seen in Fig.~\ref{fig:both_trees}.  The left hand trees correspond to the Optical dataset and the right hand ones to the Infrared dataset. Same stars are connected through lines.  The top trees are a subset of the trees from Fig.~\ref{fig:tree} and Fig.~\ref{fig:hipothesis1_monolithic_ir}, that is,  they are built using all abundance ratios but only for the 16 stars of Tab.~\ref{tab:sipsop}. Rooting the tree by the star with lowest [Ca/H], which is one of the elements with best agreement between both datasets, allows us to use the same root for both trees. But the topology does not agree among both datasets, as seen by the various crossed lines. This suggests the choice of abundances ratios used affects the order of the tips in the tree. This conflict of branches thus causes the {\it GC/Block} blue and the GC red branches of the Infrared tree to be placed at locations that are not expected from the Optical tree.

To quantify the difference, we calculate the normalized Robinson-Foulds distances ($\mathrm{RFD}$), following \cite{DaniTree}.  $\mathrm{RFD} = 0$  means trees have identical branching pattern, and $\mathrm{RFD}=1$ means trees have totally different branching patterns. An extensive explanation of this distance measure and its pros and cons in using it can be found in \cite{yang2014molecular}. Between the two upper trees of Fig.~\ref{fig:both_trees}, we obtain $\mathrm{RFD} = 0.63$.  The lower panels of Fig.~\ref{fig:both_trees} are trees built using only the common abundance between the two (see the elements plotted in Fig.~\ref{fig:chemical_comp}). The ranking of the tips between both trees agrees better, but the trees still differ by $\mathrm{RFD} = 0.57$. There are only a few stars that are strongly disordered, with 38\% of the stars ordered identically. And, if we remove SOP\_29/SIR\_58, that increases to 60\%. This seems like a large improvement on the all-abundance tree comparison, but the RFD score remains quite similar. This might be due to the uncertainties of the abundance measurements, which are driving significant scatter in the tree shapes, as noted by the averaged 35\% of node support discussed above. However, we need to consider the limitations of the RFD distance metric, which only focuses on the topology of the tree and counts the number of common nodal position between trees \citep[see][]{yang2014molecular}.  We discuss this issue further below. 

It is actually interesting that these $\mathrm{RFD}$s are not significantly different when using the full abundance dataset or only the common abundances. In order to assess if this similarity is significant given the uncertainties of the measurements, we can estimate the minimum expected difference for the same tree, given its uncertainties. To do so, we run another Monte Carlo simulation to calculate random distance matrices considering the abundances and the reported uncertainty. We then build NJ trees and compare them.  We do this 1000 times and estimate the mean $\mathrm{RFD}$ and its standard deviation for the Optical and for the Infrared sample independently.  This was done using the common abundances between two samples, as well as the full set of abundances. 

For the common set of abundances, we obtain that, given the uncertainties, the Optical samples are consistent among each other with a mean $\mathrm{RFD} = 0.42 \pm 0.11$. The Infrared samples are consistent among each other with a mean of $\mathrm{RFD} = 0.39 \pm 0.1$. Thus, the value of $\mathrm{RFD} = 0.57$ for the difference of the Optical versus Infrared tree computed using the common set of abundances, which is displayed in the lower panel of Fig.~\ref{fig:both_trees}, can be attributed to the systematic differences of the abundances of each dataset and not only to the random uncertainties of the data. 

For the full set of abundances, that is Fe, O, Na, Al, Si, Ca, Sc, Ti, Ni, La, and Eu for the Optical and Fe, O, Mg, Al, Si, Ca, and Ce for the Infrared, we obtain that trees are similar among each other on average by $\mathrm{RFD} = 0.31 \pm 0.11$ and $\mathrm{RFD}= 0.35 \pm 0.1$, respectively. These numbers are similar to each other, and comparable to the 35\% of the nodal support. 
In both cases, adding abundances increases the robustness of the tree topology accounting for abundance measurement errors, since $\mathrm{RFD}$ decreases. We note that the improvement of the  Infrared $\mathrm{RFD}$ is very small. This is because only two additional elements, namely Mg and Ce, have been added to the distance matrix. The value of $\mathrm{RFD} = 0.63$ for the difference of the Optical and the Infrared tree using all the abundances displayed in the upper panel of Fig.~\ref{fig:both_trees} is also the result of more precise trees and therefore the difference can not be explained only due to errors.

It is worth referring to to the discussion of \cite{KuhnerTree}. The RFD metric counts the number of branch partitions that appear in one tree but not the other, scoring one for each non-matched partition. This means that the $\mathrm{RFD}$ metric does not consider differences in branch lengths, only tree topology. Furthermore, a rooted-tree has a maximum of number of nodes of $2 (n - 2)$,   where $n$ is the number of tips. This means $\mathrm{RFD}$ has $2 (n - 2)$ partitions to score. In our case, $n=16$ implies we score 28 partitions only, which is not a very large number to quantify significant differences between the trees here.

Even considering these caveats, we find that the Optical and Infrared trees are different to each other. This difference is dominated by the systematic uncertainties of the measurements rather than the election of elements used in the distance matrix. 
 A reason for this result could be that even if the full Optical and Infrared sample consider different elements,  these elements overlap in three nucleosynthetic channels, namely, CCSNe, SNe-Ia, and  AGB  winds. In both datasets $\alpha-$capture, iron-peak, and neutron-capture elements are included. We note that the Optical sample includes Eu, which is an r-process element produced by neutron star mergers, but Eu has a weaker contribution to the variance of the Optical data, according to the PCA (see Sect.~\ref{sect:pca}).  Indeed, by removing for instance Fe from the Infrared sample, the $\mathrm{RFD}$ between the Optical and the Infrared sample increases to a mean of $\mathrm{RFD} = 0.92 \pm 0.01$. Removing Ce has a much smaller impact on the $\mathrm{RFD}$ estimate. This could be related to the fact that Fe is a more significant driver of the variance in the abundance space compared to Ce, as discussed in Sect.~\ref{sect:pca}. 
 
 Branching patterns, phylogenetic analyses and history reconstructions are thus highly dependent on the elements employed, but most importantly, on the abundance accuracy and precision estimates. Better interpretation on this kind of analysis will be possible once abundances reach higher level of accuracy and precision, or once simulations reach the numerical precision to model individual stars and abundances in objects like \ocen.

\subsection{Connecting branches and populations through common stars}\label{sect:common_stars_tree}

The interpretations regarding the nature of the populations found in the Optical sample depend on the dataset and the selection of stars, but using a different dataset does not contradict our results. We can reach that conclusion by studying the groups of stars in common. The trees in Fig.~\ref{fig:both_trees} show that purple stars in the Optical typically appear alongside yellow and red stars in the Infrared, while orange stars in the Optical are either yellow or blue infrared. Looking at the positions of these stars in Fig.~\ref{fig:tree} and \ref{fig:hipothesis1_monolithic_ir} allows us to conclude that the orange-purple Optical branch, as well as the yellow-red-magenta branch in the Infrared, correspond to {\it Block-Cen}, where stars show signatures of in-situ star formation. 

{\it GC-Cen} stars, dominated mainly by orange stars in the Optical, are divided into blue and red stars in the Infrared. The GMM model prefers two groups for this population, which are divided into two branches in the Infrared tree. While the metallicity range is comparable in these branches, the O-Al anticorrelation is quite different.  Indeed, among the three stars in the bottom red branch of the Infrared sample which are included in the Optical sample, SIR\_61 is colored purple, while SIR\_19 and SIR\_45 are colored gray. These three stars are actually in the {\it Child-Cen} branch in the Optical tree, but SIR\_45 is mixed with orange, and SIR\_61 is at the bottom of the gray branch. We could thus attribute the red branch to {\it Child-Cen} instead of {\it GC-Cen}, but the metallicity trend and distribution are quite narrow. 
The Infrared Stellar population bar plot shows that the blue, magenta and yellow groups are more related to each other than the red, suggesting that indeed the red is a more independent population, such as  GCs falling into the potential well of the NSC.  

It might be possible that the Infrared sample has other stars belonging to {\it Child-Cen}. We need to consider that we do not have the same information for both trees. We have abundances of different species, which are measured using various methods and data. Furthermore, we have only 20\% of stars that both samples have in common. If the Infrared dataset includes Child-Cen, it does not cover a large range in s-process and metallicity to be assigned to a different "caterpillar" branch or a new GMM group. It would be worthwhile to see if some of the most metal-rich stars observed in \cite{All_apo} indeed follow the chemical trends of the Optical sample. To do so, Optical spectroscopic follow-up observations are needed.

It is worth commenting on the recent discussions about the anticorrelations of \ocen, which tend to become weaker with higher metallicity  \citep{AlvarezGaray24, All_apo}. 
The {\it Child-Cen} anticorrelation we observe for the highest metallicity stars could be the consequence of a channel that is different from the MgAl nucleosynthetic chain of globular clusters \citep{AlvarezGaray24}. Indeed, \cite{Romano23} studied various chemical evolution models to understand the chemical pattern of Terzan 5, which was found to host multiple populations similar to \ocen, and is in the bulge of the Milky Way. In such a dense environment, the [Fe/H] spread, the age spread, and the chemical pattern of Terzan 5 is used to conclude that Terzan 5 can have multiple populations only because of in-situ star formation. Through a combination of star formation bursts with different duration and efficiency (which is dependent on the temperature of the gas, which might further depend on the temperature of the surroundings) and the ability to retain remnant stars for later Fe-peak production from SNe-Ia, it is possible to have an impact on the relative abundance ratios in such systems.  

Under the hypothesis that \ocen\ is a NSC, we are able to identify both its main formation channels, namely GC accretion and in-situ formation \citep{Neumayer20}. Using Fig.~6 of \cite{Fahrion21}, which relates the formation channels and the NSC progenitor galaxy's mass, we conclude that the progenitor galaxy of \ocen\ could have a mass of about $10^9 \mathrm{M}_\odot$. This mass has been measured for the galaxy of the largest major merger event \citep{Das+2020}, namely  Gaia-Enceladus Sausage  \citep[GES][]{helmi18, belokurov18}. Our analysis thus provides evidence to support previous claims of \cite{BekkiFreeman03} and \cite{PfefferB21}, that \ocen\ might be the NSC of the GES.  Indeed, among all clusters that are dynamically associated with the GES, according to \cite{massari19}, \ocen\ is the one with the highest binding energy \citep[see also the extensive discussion from][about this scenario]{limberg22}.

 We note that the connection between the GES and \ocen\ does not need to imply a chemical similarity between the \ocen\ populations and the field accreted stars that have been associated with GES by various works in the literature. Indeed, these stars are more $\alpha-$poor, Al-poor, and Eu-rich compared to \ocen\ \citep[see also Fig.~\ref{fig:chemtrends_gmm} to see some elements][]{Matsuno21, carrillo22, 2011A&A...530A..15N, Das+2020, dasilva23}. The chemical similarity between GES and \ocen\ might be expected if \ocen\ were a typical GC, such as NGC1261, which \cite{massari19} have attributed be part of GES. Indeed, \cite{Koch21} provided evidence of its chemical similarity with GES. But expecting stars from a NSC and the field of the parent galaxy sharing the same chemical composition might imply they share the same star formation history, which does not need to be true \citep{Neumayer20}. 

 Let us furthermore consider the current belief that M54 is the NSC of the Sagittarius (Sgr) galaxy \citep{alfaro-cuello19}.  The chemical abundance analysis of M54 and the Sgr performed by \cite{carretta10} showed a continuation of $\alpha$ elements between M54 and the Sgr core, but the metallicity of both populations did not overlap. Abundances such as O, Na, Al differ due to the anticorrelations present in M54, which are not in the field. In \cite{2024arXiv241206896V}, more metal-poor stars were included in the analysis of the field stars, overlapping better with the M54 metallicity coverage of \cite{carretta10}. The overlap shows that $\alpha-$ abundances of Sgr's core at a $\mathrm{[Fe/H]} \sim -1.5$ is of the order of 0.2 dex, while the abundances of M54 is of the order of $0.4-0.5$ dex, although this comparison is subject to uncertainties due to the different methods employed in these independent works. This difference is consistent with what is found between GES and \ocen.  More discussions about the chemical connection between GES stars and \ocen\ and the role of europium can be further found in App.~\ref{sect:europ}.

\subsection{The history of the formation and evolution of \ocen}

We assume that NSCs are dense stellar systems, which contain stellar populations formed in different ways. Under the hypothesis that \ocen\ is a NSC, whose host galaxy has been disrupted by its accretion on to the Milky Way, we can use the various populations of \ocen\ to understand not only its formation history, but also the properties of \ocen's parent galaxy. 

Our phylogenetic analysis using NJ trees allows us to study how our GMM groups are distributed and related to each other, enabling us to deduce that \ocen\ appears to have formed via multiple formation channels: accretion of globular clusters, accretion of gas and stars belonging to primordial building blocks of galaxies, as well as in-situ star formation. GC-like stars belonging to G\_O1-orange, are the most metal-poor and probably the most ancient ones in our dataset. Furthermore, the [La/Fe] abundances of these stars are low compared to the rest of the sample, and have a narrow distribution as well. This is consistent with a population formed rapidly and early on. They might be the relics of GC accretion onto \ocen. Because {\it GC-Cen} is mono-metallic, with a rather small dispersion, our data does not support a contribution of a large number of GCs contributing to the formation of this NSC.  The other stellar populations in \ocen\ show signatures of being the result of prolonged star formation. These populations belong mostly to G\_O2-gray and G\_O3-purple but some belong to G\_O1-orange and are separated in two `caterpillar' branches in the NJ tree.

The metallicity distribution of {\it Block-Cen}, as well as its [O/Fe] and [La/Fe] trends, are consistent with chemical evolution models of \cite{Romano23}, which are specifically designed to replicate abundance patterns in dense and old environments, in their case, Terzan 5. Among the various abundance ratios, the [Al/Fe] enhancement  has been found in potential ancient building blocks of the Milky Way \citep{MalhanRix24, Buckley24}. The fact that {\it Block-Cen} is enhanced in [Al/Fe] might thus be the result of a primordial population formed at the center of \ocen's progenitor galaxy, which was accreted early on to form the NSC along with the accreted globular clusters. It is possible that both stars and gas were accreted onto \ocen. 

The second population, {\it Child-Cen}, corresponds to the branch with gray stars.   Its [Fe/H] as well as [La/Fe] reach values that are very enhanced. [Al/Fe] is also very enhanced in this population, whereas the [O/Fe] values are very depleted. The low [O/Fe] might be a signature of pristine gas accretion onto \ocen\ \citep{dercole12}.  While a combination of star formation bursts and interactions with the surrounding gas of a dense environment might yield an abundance pattern like the one found here, the pattern of a population like {\it Child-Cen} has not been intentionally modeled so far.  The abundance trends of this branch are not like those observed in globular clusters,  dwarf galaxies or in the populations of the Milky Way as discussed above, therefore we attribute it to a population that formed in situ. 
Our phylogenetic analysis has allowed us to discover abundance trends that would require new models to explain them.  

 As a final note, we point to the recent independent analysis of \cite{mason25}, which was announced at the same time as this work. That work used a different dataset and different methods to reach a remarkably consistent conclusions about the three populations of \ocen: P1 (our {\it Block-Cen}), P2 (our {\it Child-Cen}), and IM (our {\it GC-Cen}). While the similarities and differences of the exact properties and abundance patterns of these populations need to be further studied, it seems that new data and methods are converging into disentangling the populations of \ocen\ and revealing its nature as a NSC.

\section{Concluding remarks}\label{sect:concl}

Our interdisciplinary methodology is able to provide more sophisticated analyses of the astrophysical evolution of a complex stellar system such as \ocen\ than has been viable using traditional, purely astronomy approaches. Our interpretation that \ocen\ might have built up from the merging of at least one metal-poor GC and more metal-rich populations whose broad metallicity distribution could be the result of prolonged star formation agrees well with the recent conclusions of \cite{AlvarezGaray24} and the overall recent literature about the theoretical connection between \ocen\ and NSCs \citep{2018Brown, PfefferB21, Gray24}. \cite{AlvarezGaray24} claim to reach their conclusions because of the improved precision in their abundance measurements compared to \cite{All_ctio}. Here, however, we use these seemingly less precise datasets, but are still able to reach similar conclusions to  \cite{AlvarezGaray24}. 
This is thanks to the  novel methods we employ, adapted from  evolutionary biology, which has an established history in performing population analysis reconstructions

Nuclear star clusters might result from GC mergers, accretion of ancient building blocks, and in-situ star formation \citep{Neumayer20}. Our analysis presents new evidence in favor of\ocen\ being a nuclear star cluster. We find that its stellar populations have different chemical abundance pathways and relationships. The chemical pattern of the various branches in our NJ trees suggest that, in contrast to regular galaxies or genuine globular clusters, the \ocen\ chemical history has been impacted by other processes, such as strong accretion and mixing. The multiple-channel formation scenario of NSCs can indeed account for the large variety of abundance patterns observed in \ocen. Our phylogenetic analysis enables us to navigate the multidimensional chemical dataset produced by these formation pathways and disentangle the contributions of each channel. Furthermore, our analysis allows us to connect all these populations and reconstruct a possible shared history.  

The presence of more than one `caterpillar' branch in our tree thus rejects the hypothesis that \ocen\ had an isolated evolutionary history.  In fact, our branches could be associated with stellar populations of different origins: (1) {\it GC-Cen}:  a globular cluster population; (2) {\it Block-Cen}: a stellar population that experienced star formation for more than one Gyr, with similar properties and perhaps origins to ancient clumpy populations found in dense environments such as the center of our Milky Way.  
(3) {\it Child-Cen}: a population with in-situ star formation, which could have formed inside \ocen, but its chemical pattern has not been seen before so explaining it requires tailored modeling. Whether these are the true and only origins of the stellar populations of \ocen, however, remains to be seen with more data, high resolution simulations of NSCs, or with more detailed spectroscopic analysis of other NSCs.   

With this work, we  demonstrate the large prospects of applying phylogenetics to studying the historical processes in our cosmic surroundings. As long as heritable markers are present in our data, phylogenetic methods can be used.
As we come to appreciate the full complexity of the evolution of the cosmic bodies such as the emblematic Omega Centauri, we must be willing to expand our horizons, even beyond astronomy, to further advance our understanding.

\section{Data availability}

Tables with abundance determination as introduced in Sect.~\ref{sect:data} are only available in electronic form at the CDS via anonymous ftp to \url{cdsarc.u-strasbg.fr} (130.79.128.5) or via \url{http://cdsweb.u-strasbg.fr/cgi-bin/qcat?J/A+A/}.

\begin{acknowledgements}
This paper accomplishes one of the main scientific objectives of the interdisciplinary project Millennium Nucleus for the Evolutionary Reconstruction of the InterStellar medium (ERIS NCN2021\_017). This work has been supported through student thesis fellowships of P.S, and K.W, in addition to funding for travel for P.J and K.W to visit or invite P.D, R.Y and X.H in several occasions. P.J. and C.A.G. thank Keaghan Yaxley, Sven Buder, Joshua Povick and Nicole Buckley for fruitful suggestions and exploratory analyses on this paper. P.J. and F.C warmly thank Macarena Concha for introducing us to each other. P.J furthermore thanks FONDECYT REGULAR  1231057. C.A.G. acknowledges support from  FONDECYT Iniciación 11230741. P.V and P.C acknowledge Milenio – ICN17\_022 and NCN2024\_040. P.D is supported by a UKRI Future Leaders Fellowship (grant reference MR/S032223/1). S.V. thanks ANID Beca Doctorado Nacional, 21220489). P.B.T acknowledges FONDECYT Regular 1240465.  P.B.T and E.J.  acknowledge support from ANID Basal Project FB210003. This project has received funding from the European Union Horizon 2020 Research and Innovation Programme under the Marie Sklodowska-Curie grant agreement No 734374-LACEGAL. A.R.A. acknowledges support from DICYT through grant 062319RA  FONDECYT Regular  1230731. GG acknowledges support from the Leverhulme Trust under grant EM-2025-007. The authors warmly thank the referee for the constructive and respectful feedback. 
\end{acknowledgements}

\bibliographystyle{aa} 
\bibliography{references2}

\begin{appendix}
\onecolumn

\section{Stars in common between Optical and Infrared sample}
Table~\ref{tab:sipsop} lists the stars in common between both samples. The identification number from Gaia DR3, the APOGEE ID and the Leid ID are indicated for these stars, in addition to our own designation. We add the Gaia G apparent magnitude for reference. 

\begin{table*}[h!]
    \centering
\begin{tabular}{rlllrr}
\hline
  \multicolumn{1}{c}{SOURCE\_ID} &
  \multicolumn{1}{c}{StarID\_IR} &
  \multicolumn{1}{c}{StarID\_Opt} &
  \multicolumn{1}{c}{APOGEE\_ID} &
  \multicolumn{1}{c}{LEID\_ID} &
  \multicolumn{1}{c}{Gmag} \\
\hline
  6083709162042645504 & SIR\_019 & SOP\_018 & 2M13265592-4722213 & 29069 & 11.877\\
  6083506439583009920 & SIR\_035 & SOP\_122 & 2M13263895-4743584 & 76027 & 12.012\\
  6083516197752198144 & SIR\_045 & SOP\_092 & 2M13273759-4730522 & 48409 & 12.445\\
  6083725379834365952 & SIR\_048 & SOP\_002 & 2M13264112-4714262 & 11021 & 12.234\\
  6083705386799464576 & SIR\_058 & SOP\_029 & 2M13272390-4724345 & 34207 & 11.838\\
  6083740979155658368 & SIR\_061 & SOP\_026 & 2M13244676-4724486 & 34008 & 12.350\\
  6083704008080853632 & SIR\_064 & SOP\_068 & 2M13273206-4728228 & 42497 & 12.281\\
  6083713633137662208 & SIR\_083 & SOP\_057 & 2M13252908-4727205 & 40016 & 12.196\\
  6083703084671715328 & SIR\_086 & SOP\_023 & 2M13265189-4723468 & 32101 & 12.634\\
  6083699438234822400 & SIR\_098 & SOP\_108 & 2M13260041-4734308 & 55028 & 11.852\\
  6083509295752195200 & SIR\_102 & SOP\_119 & 2M13265181-4736480 & 61070 & 11.912\\
  6083701156227097088 & SIR\_106 & SOP\_063 & 2M13261171-4728283 & 42054 & 11.093\\
  6083515544917590528 & SIR\_107 & SOP\_105 & 2M13273242-4733078 & 53185 & 11.354\\
  6083708303049198336 & SIR\_116 & SOP\_017 & 2M13271156-4722009 & 28084 & 12.570\\
  6083514273605994112 & SIR\_125 & SOP\_121 & 2M13272837-4738131 & 64067 & 11.938\\
  6083714629570589696 & SIR\_126 & SOP\_022 & 2M13262962-4723413 & 32063 & 12.294\\
\hline\end{tabular}
    \caption{IDs and Gaia magnitude of stars in common between both samples}
    \label{tab:sipsop}
\end{table*}

\FloatBarrier 
\twocolumn

\section{Principal component analysis and Gaussian mixture models}

\subsection{PCA}\label{sect:pca}

\begin{figure}[h!]
\centering
\includegraphics[scale=0.45]{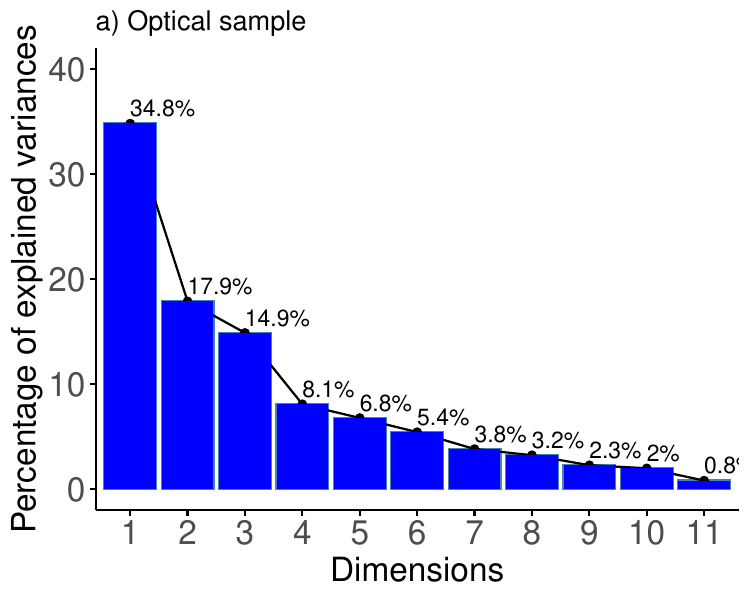}
\includegraphics[scale=0.45]{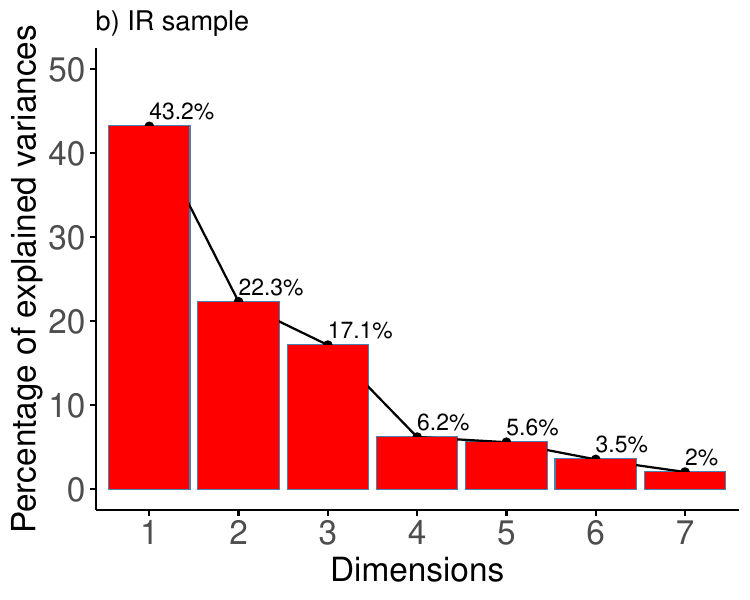}
\caption{Contributions to the chemical variance of each PCA component for the Optical (a) and Infrared (b) samples.    }
\label{fig:pca_contrib}
\end{figure}

Figure~\ref{fig:pca_contrib} illustrates the PCA outcomes, with the Optical sample displayed in blue on the left and the Infrared sample in red on the right.  The histograms show the percentage of the variances of each principal component (dimension).  We first focus on the Optical sample, which has abundance ratios of 11 elements.   While the first three components explain 67\% of the variance, with the second and the third dimensions contributing to a similar percentage (between 15 and 18\%), a total of six dimensions are needed to explain 87\% of the variance in the chemical space.

The situation is slightly different regarding the Infrared sample, which has fewer abundances measured than the Optical sample. The first two components already explain more than 60\% of the variance, and the first three components explain 80\% of it. We comment that even though the Infrared sample is more limited in several elements, the sample also includes iron-peak, $\alpha-$capture, and neutron-capture elements. However, it lacks the heavy r-process neutron-capture elements.  The percentages cannot be directly compared to the Optical sample, since they are normalized to their own dataset.

\begin{figure*}[h!]
\centering
\includegraphics[scale=0.45]{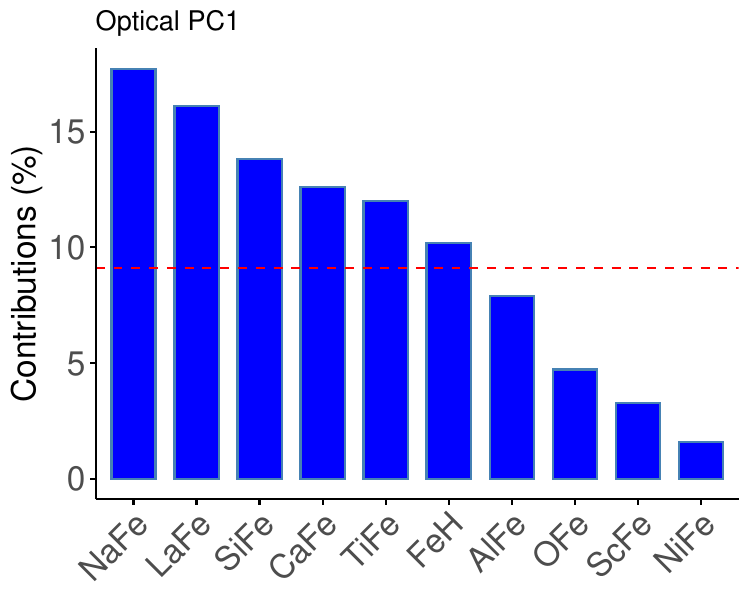}
\includegraphics[scale=0.45]{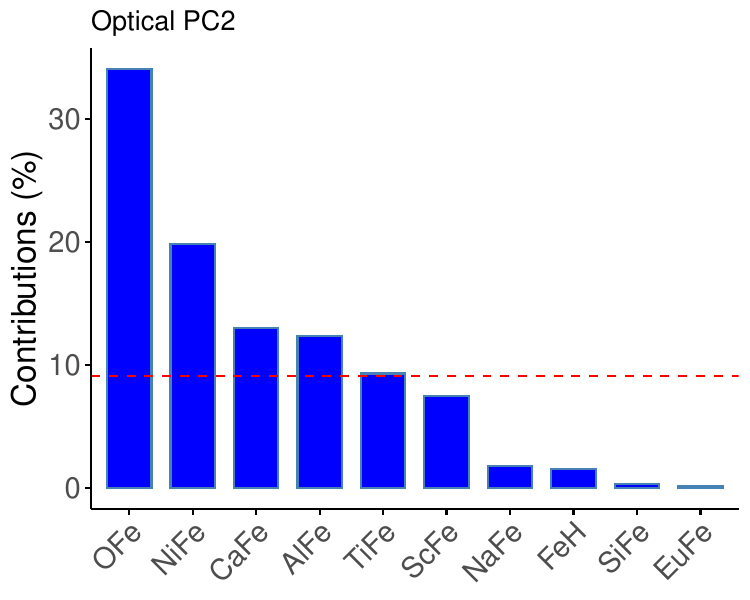}
\includegraphics[scale=0.45]{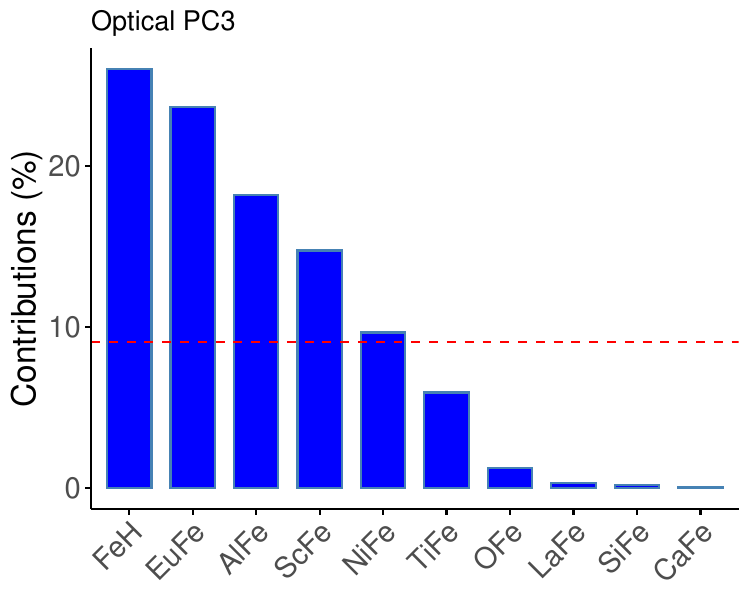}
\includegraphics[scale=0.45]{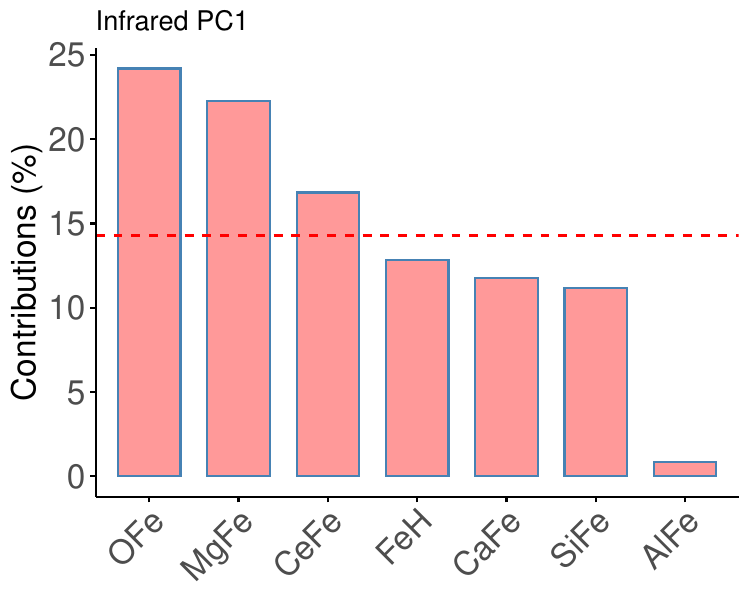}
\includegraphics[scale=0.45]{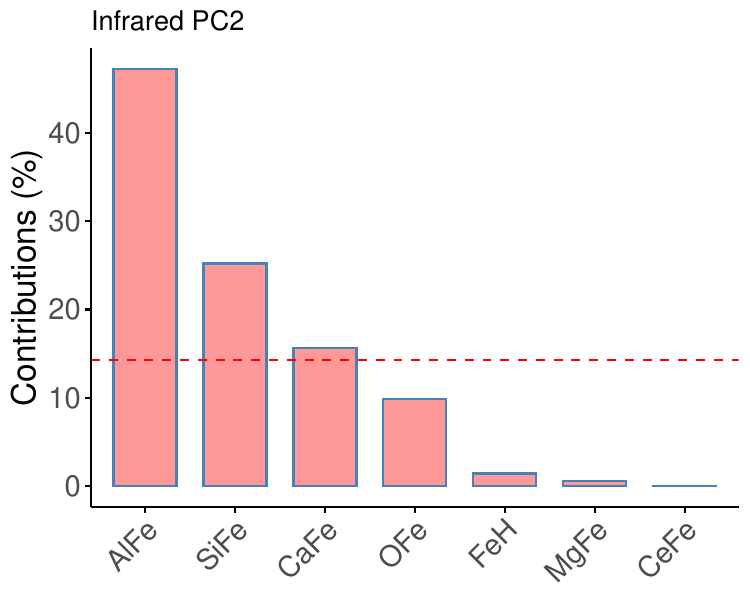}
\includegraphics[scale=0.45]{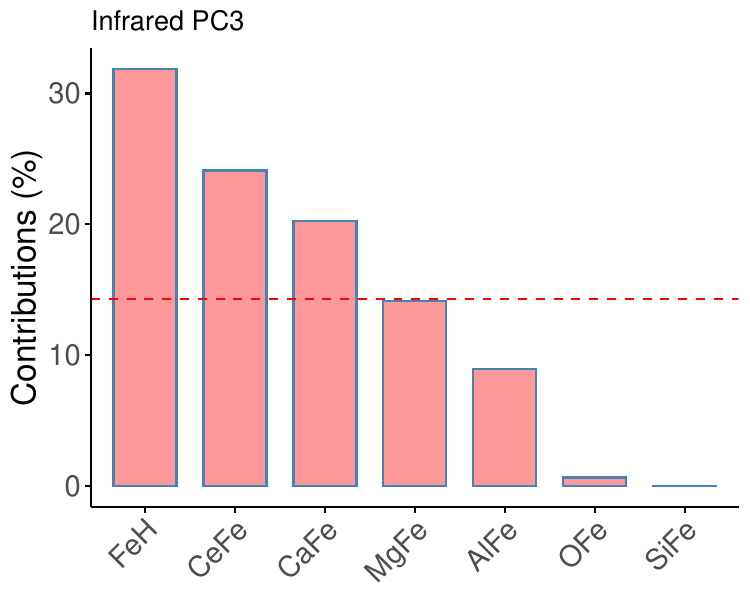}
\caption{Contributions to the first principal components from each element. In blue the Optical Sample, in red the Infrared sample. The dashed red horizontal line delineates the 50\% of contribution, helping us to see which elements contribute most each PC. }
\label{fig:pca_contributions_2}
\end{figure*}

Figure~\ref{fig:pca_contributions_2} shows the contribution of the individual elements in each principal component, with the dashed horizontal line indicating the threshold of 50\% of contribution. The upper panels show the Optical sample and the lower panels show the Infrared sample. 

For the Optical sample, we obtain that Na, La, Si, Ca, Ti, and Fe are the elements that most contribute to the first component. O, Ni, Ca, and Al contribute most to the second component, and Fe, Eu, Al, and Sc contribute to the third component. We thus find that elements coming from different nucleosynthetic paths are combined in each principal component. The Infrared sample shows a similar situation. The most influential elements in the first component are O, Mg, and Ce, in the second component are Al, Si, and Ca, and the third component involves Fe, Ce, and Ca.

\begin{figure}[h!]
\centering
\includegraphics[scale=0.6]{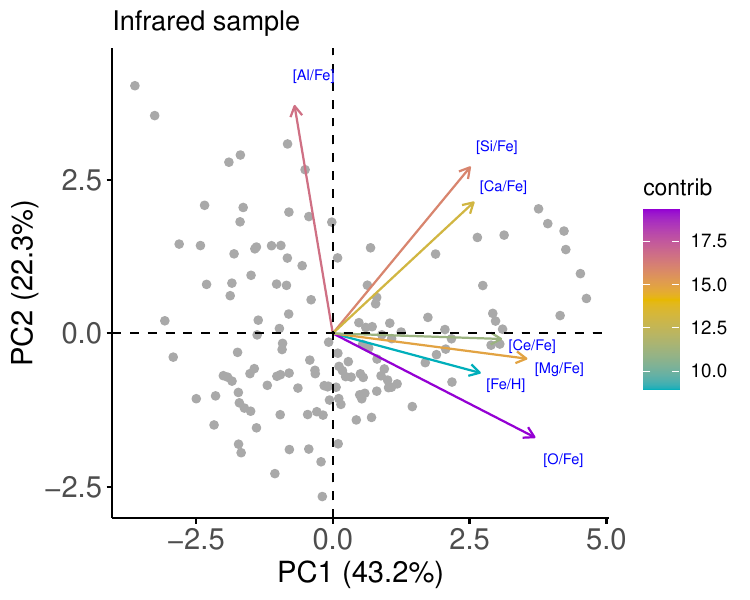}
\includegraphics[scale=0.6]{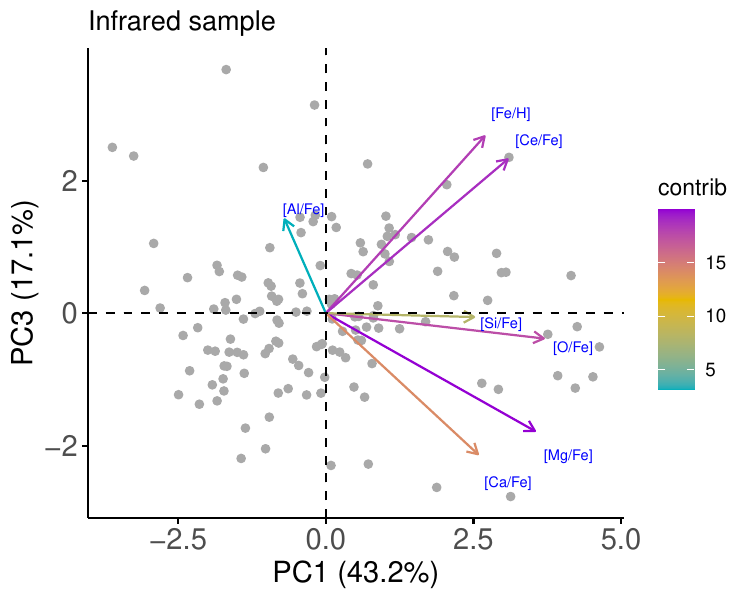}
\caption{PCA planes with the direction and contribution of each chemical element in the first three most significant components of the Infrared sample. Stars are distributed in the planes and arrows indicate the direction to which the abundance ratios are higher in these planes. Colors indicate the contribution to the PCA of the abundance ratios. The first two dimensions show how [O/Fe] and [Al/Fe] dominate the variance, contributing to opposite directions in these planes. This is due to the anticorrelation of these elements. }
\label{fig:pca_planes2}
\end{figure}

In Fig.~\ref{fig:pca_planes2} we plot the PC planes of the Infrared sample, in a similar way to Fig.~\ref{fig:pca_planes} for the Optical sample.  
The left-hand panel, focused on the first and second PC of the Infrared sample, shows a similar picture to the Optical first and second PCs, in the sense that [Al/Fe] and [O/Fe] arrows point towards opposite directions, indicating that they anticorrelate.  [O/Fe] has a long and purple arrow, showing the dominance of [O/Fe] in the variance of the Infrared sample. Another arrow that points in the same direction but is shorter than the [O/Fe] one is the [Mg/Fe] arrow. This element is responsible for  this dataset's anticorrelation with [Al/Fe]  \cite{All_apo}. The fact that Mg and O have similar direction means they correlate. This is expected, based on that they have similar trends in Fig.~\ref{fig:abund}.

The third principal component, in the right-hand panel of Fig.~\ref{fig:pca_planes2} shows the importance of the iron-peak and the neutron-capture elements in the variance, in a similar way than in the Optical case.  We further note that Ca never plays a similar role in importance to the Optical case.  Si and Ca  arrows are orthogonal to O and Al in both samples, which indicates that these $\alpha-$capture elements do not correlate with O in \ocen. The s-process element Ce, and the iron-peak element Fe, although showing a small contribution by their shorter and greener arrows, they all point in the same direction, which is to the right in these planes.  In fact, in these planes only O and Al have opposite directions from the Optical sample.

\subsection{Identification of populations with Gaussian mixture models}\label{sect:gmm} 

\begin{figure}[h!]
\centering
\includegraphics[scale=0.5]{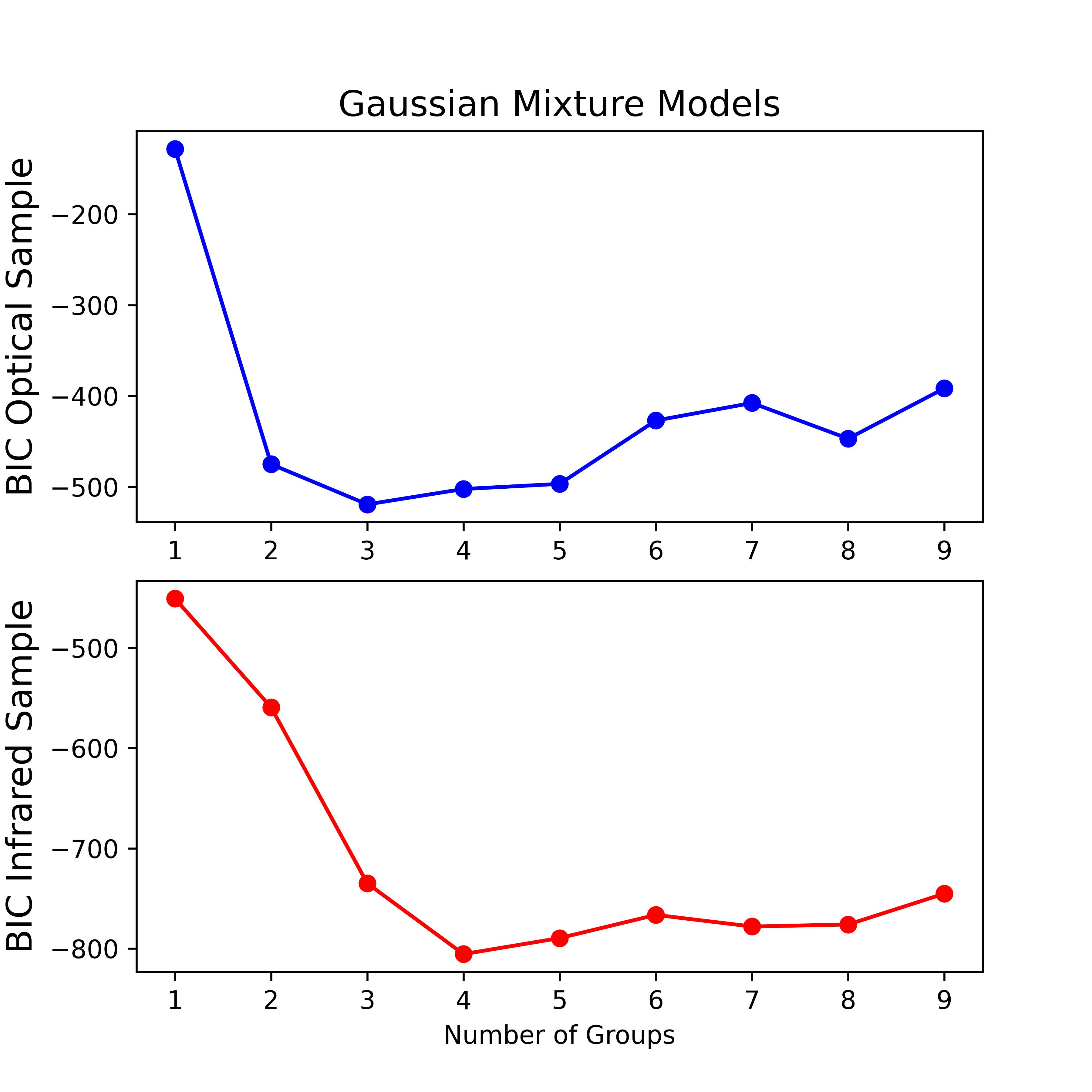}
\caption{BIC Scores for the Gaussian mixture models as a function of number of groups for the Optical sample (upper panel) and the Infrared one (lower panel). The Optical sample prefers three Gaussians and the Infrared sample prefers four Gaussians.   }
\label{fig:BIC_scores}
\end{figure}

We define populations by groups that are chemically coherent. To do so, we perform Gaussian mixture models (GMM) on the stellar abundance data considering all elements and the element space, because from the PCA exploration  we found that all elements contribute to the variance and that abundances are not so correlated with each other.  We use the {\tt scikit-learn} python package \citep{scikit-learn}  and explore the Bayesian Information Criterion (BIC) scores, which finds a tradeoff between the number of parameters and the model to fit \citep[see][for extensive discussions about GMM applied to metal-poor stars]{Buckley24}. BIC scores provide a way to perform a model selection using a penalty proportional to the number of estimated parameters, which help us avoid overﬁtted solutions. The minimum BIC scores represent the optimal combination of estimated parameters in the GMM model. We thus calculate the BIC scores for different number of groups considering each Gaussian has its own diagonal covariance matrix. The BIC scores are plotted in Fig.~\ref{fig:BIC_scores}. The upper panel shows the Optical sample in blue, and the lower panel shows the Infrared sample in red. 

The Optical sample supports three groups, which corresponds to the lowest BIC scores. The Infrared sample supports four groups. 
The populations found from the metallicity distribution of \ocen\ by \cite{All_apo} (using more stars from the same Infrared sample as us), \cite{sollima05}, as well as \cite{AlvarezGaray24}, yield four groups.  This number is however smaller than the five populations found from the metallicity distribution analyzed by \cite{All_ctio}, who used a larger version of our Optical sample, or the seven populations found clustering the chemical space using the entire sample of our Infrared sample by \cite{All_apo}.  This is a further argument against attempting to establish a number of populations in \ocen, because it heavily depends on the data, method, and the information used. See also further recent discussions on GMM for \ocen\ chemical abundances  in \cite{Pagnini}. In fact the GMM analysis on chemical GALAH data by \cite{Buder+2022} shows that the BIC scores and the optimal number of groups have a significant impact when varying the abundance ratios considered for the GMM.  In this paper we thus use GMM to see how the various groups might be related to each other and which abundances might be causing the separation of the groups.

\begin{figure*}[h!]
\centering
\vspace{-1cm}
\includegraphics[ scale=0.29]{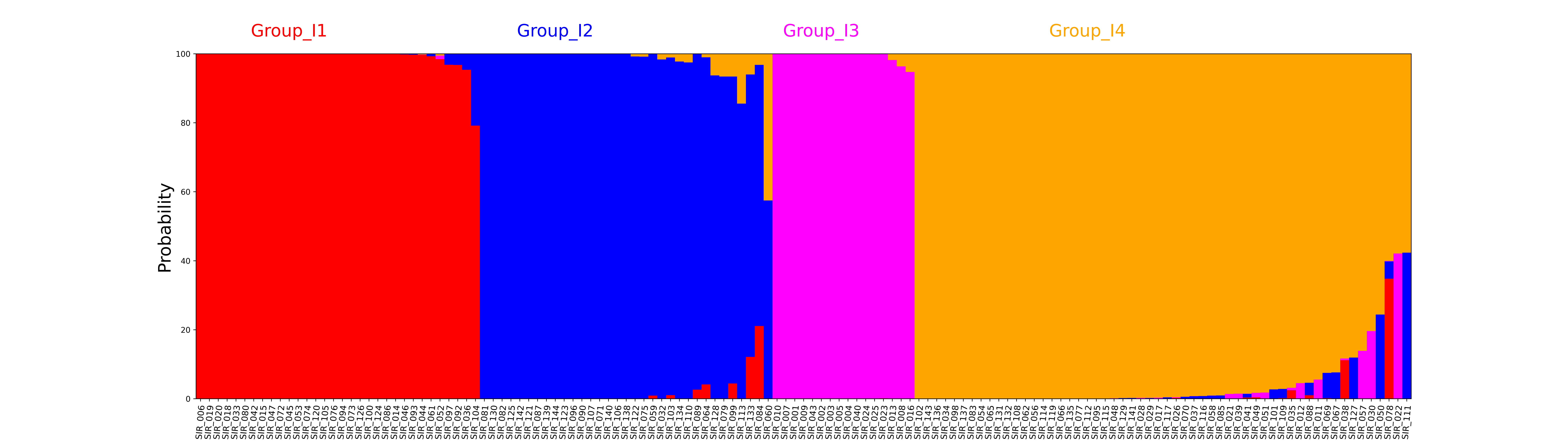}
\caption{Stellar population bar plot of the Infrared sample. Same as Fig.~\ref{fig:structure_gmm} but using four groups, which is the optimal number  for the Infrared sample from the GMM analysis. }
\label{fig:gmm_structure_ir}
\end{figure*}

Figure \ref{fig:gmm_structure_ir} shows the stellar population bar plot for the Infrared sample, with four groups colored with red, blue, magenta, and yellow. As in the case of the Optical sample, the admixture between groups is low, especially for the Group\_I1-red. A lightly higher admixture between the Group\_I2-blue, Group\_I3-magenta, and Group\_I4-yellow is seen. Following the arguments used for the Optical sample, we can interpret this stellar population bar plot by claiming that Group\_I1-red might have had a more distinct history than Group\_I2-blue, Group\_I3-magenta, and Group\_I4-yellow, which are more related to each other. 

\begin{figure*}[h!]
\centering
\includegraphics[scale=0.25]{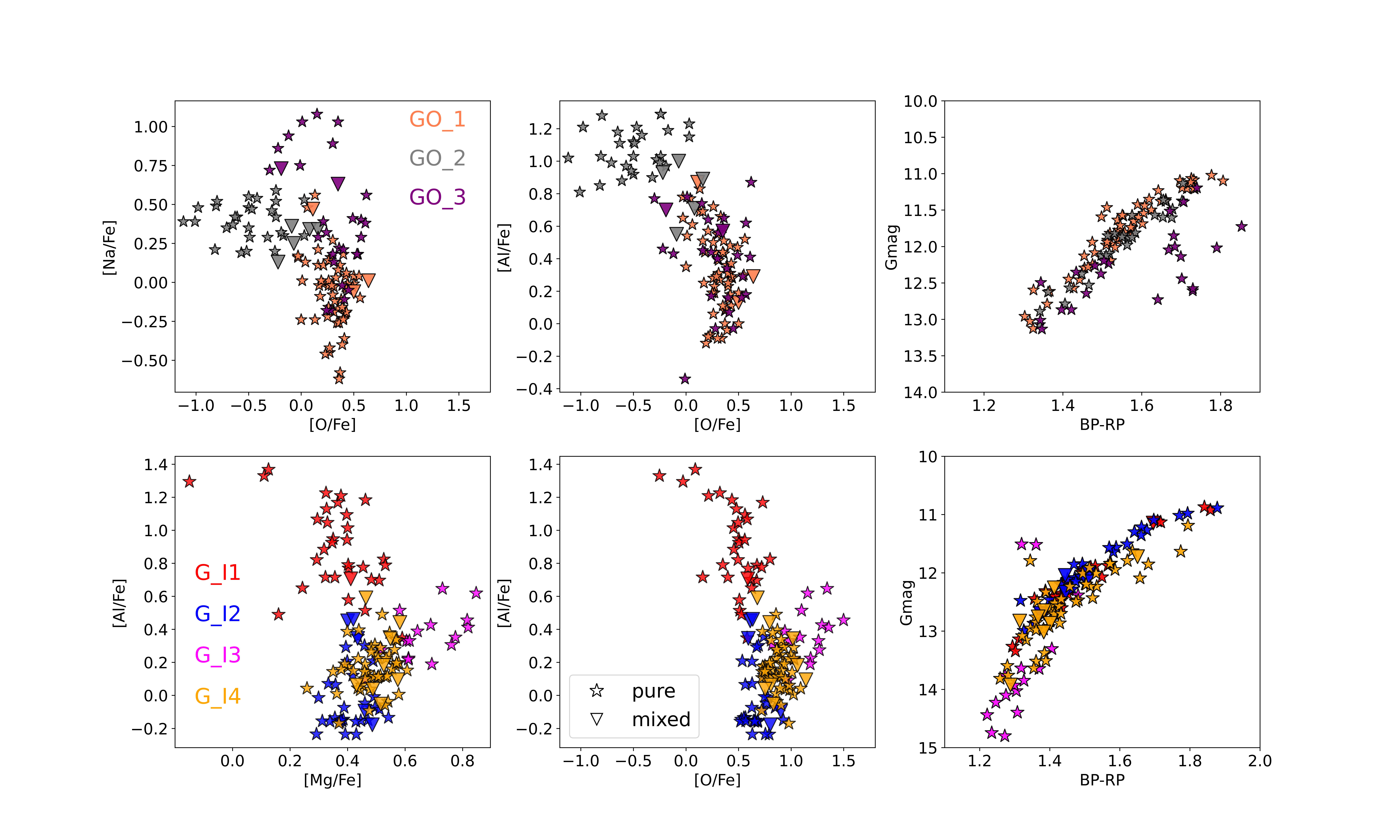}
\caption{Anticorrelations colored by groups following GMM classification and color scheme of of the GMM models. Each star is colored by the group which has the maximum probability. The right hand panels show the Color Magnitude Diagrams. Top row: optical sample. Bottom row: near infrared sample.  Shaded contours represent the KDE of the datapoints. }
\label{fig:anticorr_gmm}
\end{figure*}

\subsection{Chemical distributions of the populations}\label{sect:chemical_distr}

We present a description of the groups general properties. We focus our description on their chemical properties and their stellar parameters only. 
Regarding the kinematical properties, we investigated the proper motions from Gaia and did not find any particular evidence that some group had different kinematical properties than the other. 
Regarding the ages, we attempted to estimate ages of the individual stars, but our results could not be validated and thus we could not trust them. Acknowledging that He, C, N, and O abundances in \ocen\ are peculiar \citep{All_apo, Romano07, Tailo16} and that we do not have He abundances for our stars, we decided not to consider ages in this analysis. Indeed, \cite{Marino12} and \cite{sollima05} extensively discuss on the caveats of using ages to study \ocen\ given these element abundance peculiarities.

The anticorrelations Na-O and Al-O for the Optical sample are plotted in the top panels of Fig.~\ref{fig:anticorr_gmm}. The lower panel shows the anticorrelations Al-Mg and Al-O for the Infrared sample. The right-hand panels are the color magnitude diagrams, this allows a 1-1 comparison between the samples because the parameters come from the same source. This time the stars are colored by the population with highest frequency for each star, and symbols correspond to if the stars are pure (probability of belonging to the GMM group of more than 90\%) or mixed (probability of belonging to the group of less than 90\%). Changing this threshold does not change our interpretation, since the admixture of the stellar population bar plot is in general very low.  

The GMM finds that the populations found are well separated in the anticorrelations. This is expected since already from the PCA  we could see that the anticorrelations were dominating the variance of the data (see Fig.~\ref{fig:pca_planes} and \ref{fig:pca_planes2}). Regarding the Optical sample, G\_O2-gray stars have high [Na/Fe] and high [Al/Fe] but low [O/Fe] abundances. G\_O3-purple has very high [Na/Fe] but normal [Al/Fe] and [O/Fe]. We note they do not overlap in the red giant branch like the other stars. \cite{Pancino2000} showed that \ocen\ has multiple red giant branches, with the faintest branch being the most metal and calcium rich one.  The high [Na/Fe] ratio can be related to a metal-rich progenitor since Na production has a strong dependency of metallicity \citep{2018Brown}. 
G\_O1-orange mix in Na, Al, O, and the color-magnitude diagram, belonging to the low-Na, low-Al, and high-O side of the anticorrelation. Mixed stars are at the central parts of the diagrams.  

Regarding the Infrared sample, we see in the bottom panels of Fig.~\ref{fig:anticorr_gmm} that G\_I1-red populates the high-Al low-Mg and low-O part of the anticorrelation.  G\_I3-magenta has the highest Mg and O abundances, but Al is not particularly low. In fact, G\_I2-blue and G\_I4-yellow dominate the low-Al, high-Mg side of the anticorrelation. All these stars overlap in these diagrams, as well as in the color-magnitude diagram of the bottom right panel. As in the Optical sample, mixed stars appear at the central parts of the diagrams, where populations overlap. 

\begin{figure}[h!]
\includegraphics[scale=0.25]{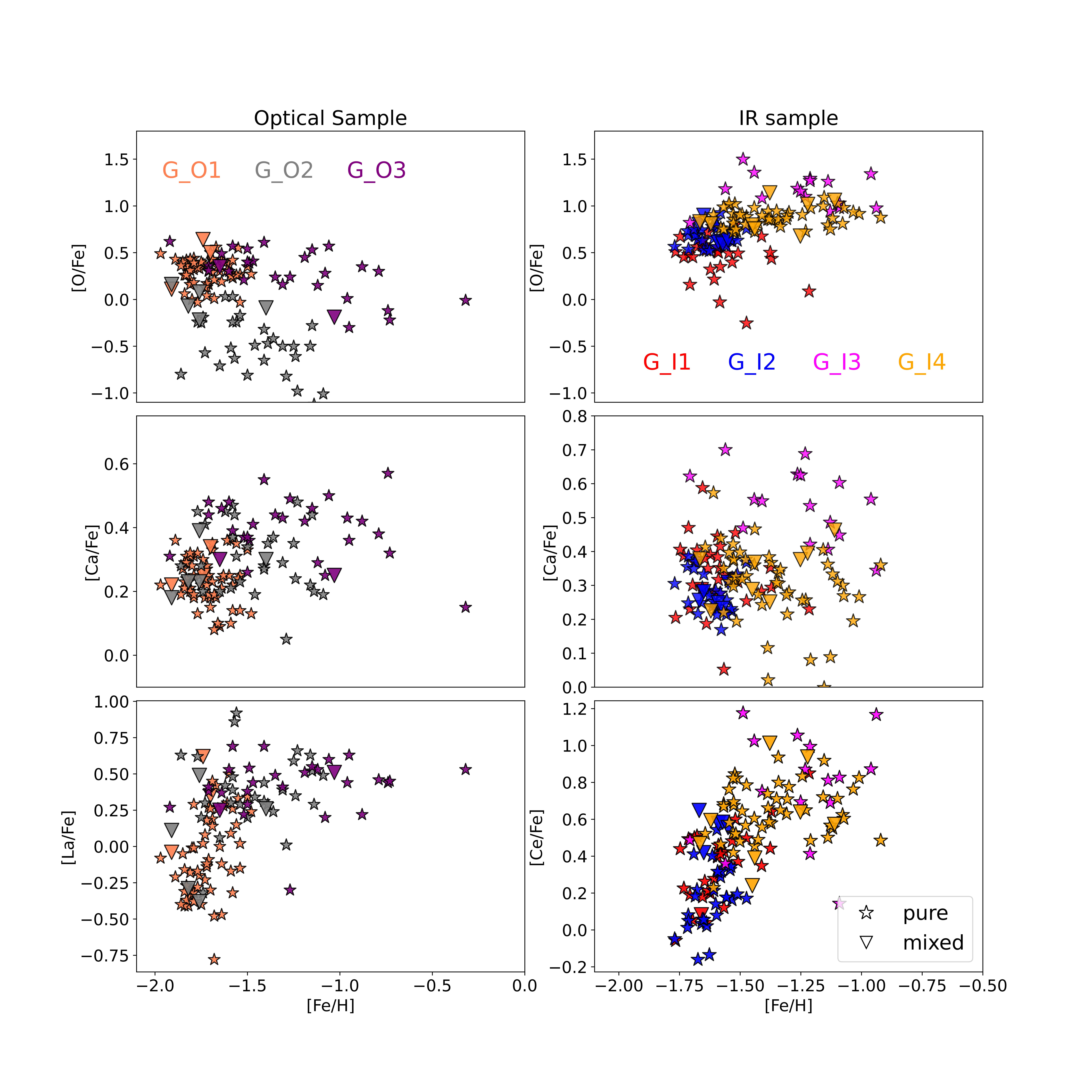}
\caption{Chemical trends colored by groups following classification and color scheme of Fig.~\ref{fig:structure_gmm} and Fig.~\ref{fig:anticorr_gmm}.}
\label{fig:chemtrends_gmm}
\end{figure}

We further explore the chemical abundance trends as a function of metallicity for our populations in Fig.~\ref{fig:chemtrends_gmm}. The Optical sample is displayed in the left-hand panels, and the Infrared sample is shown in the right-hand panels. We show the [O/Fe], [Ca/Fe], and the s-process [La/Fe] or [Ce/Fe] abundances.  

We first  realize that the populations found with our GMM method are not the same as those found from the metallicity distribution peaks by \cite{All_apo}, \cite{All_ctio} or by \cite{sollima05} and \cite{AlvarezGaray24}. While our groups separate in [Fe/H], some overlap significantly. They result in different GMMs because of the other elements considered in this analysis. Further discussions on GMM using different abundance planes can be found in \cite{Buder+2022} and how to separate groups better using GMM with PCA can be found in \cite{Buckley24}.

In the diagrams of Fig.~\ref{fig:chemtrends_gmm} we find that G\_O3-purple is the most metal-rich population of the Optical sample. The correlation between Na and Fe is thus expected \citep{2018Brown}, as well as the position of the magenta stars in the color-magnitude diagram \citep[see also][]{Pancino2000}. Both G\_O2 and  G\_O3 have high  [Ca/Fe] and  [La/Fe] abundances. \cite{Pancino2000} also points to the [Ca/H]-rich stars being Ba enriched due to AGB pollution. This is consistent with the mangenta stars having high levels of enhancement in [La/Fe]. If intermediate-mass AGB pollution is responsible for these enhancements, then some timespan of at least 0.5 Gyr is needed for this enrichment to happen. This suggests that these a stars might have formed at a later time compared to the G\_O1 stars. 

It is quite interesting to note that there are G\_O3-gray stars that are also metal-rich and [Na/Fe] rich, but they follow the red giant branch of the CMD diagram,  are considerably bluer and brighter compared to the purple stars. This can be explained by their difference in oxygen, but this could be also a result of ages, with the gray stars being younger than the purple ones. We are unable to determine individual ages to quantify this difference, given the strong chemical peculiarities of these stars.    

The G\_I3-magenta stars of the Infrared sample are the most O, Ca, and Ce-rich ones, and span a wide range of [Fe/H]. Their location in the color-magnitude diagram is mixed with the rest of the stars, but this could be a selection effect of stellar parameters between these and the Optical sample. G\_I4-yellow stars have middle Ca and Ce abundances. G\_I1-red, on the other hand, while being those with the highest Al abundances do not have significantly high Ce abundances. 
G\_I2-blue have similar properties to  G\_O1-orange from the Optical sample, namely low metallicities, O and Ca abundances of the order of 0.4 dex, and low Ce s-process abundances. 

 We plot in the lowest left panel of Fig.~\ref{fig:chemtrends_gmm} the distribution of [Eu/Fe] vs [Fe/H] for the stars colocoloredred by the GMM groups. One can see that the lowest metallicity stars, regardless of the GMM group, have higher [Eu/Fe]. The decreasing trend with [Fe/H] can be noticed, but the scatter is large. This scatter, as well as negative trend with [Fe/H], explains why Eu is important in the PCA (see Fig.~\ref{fig:pca_planes}).

\FloatBarrier
\onecolumn 
\section{Infrared sample tree}\label{sect:ir_tree}

\begin{figure*}[h!]
\centering
\includegraphics[scale=0.3]{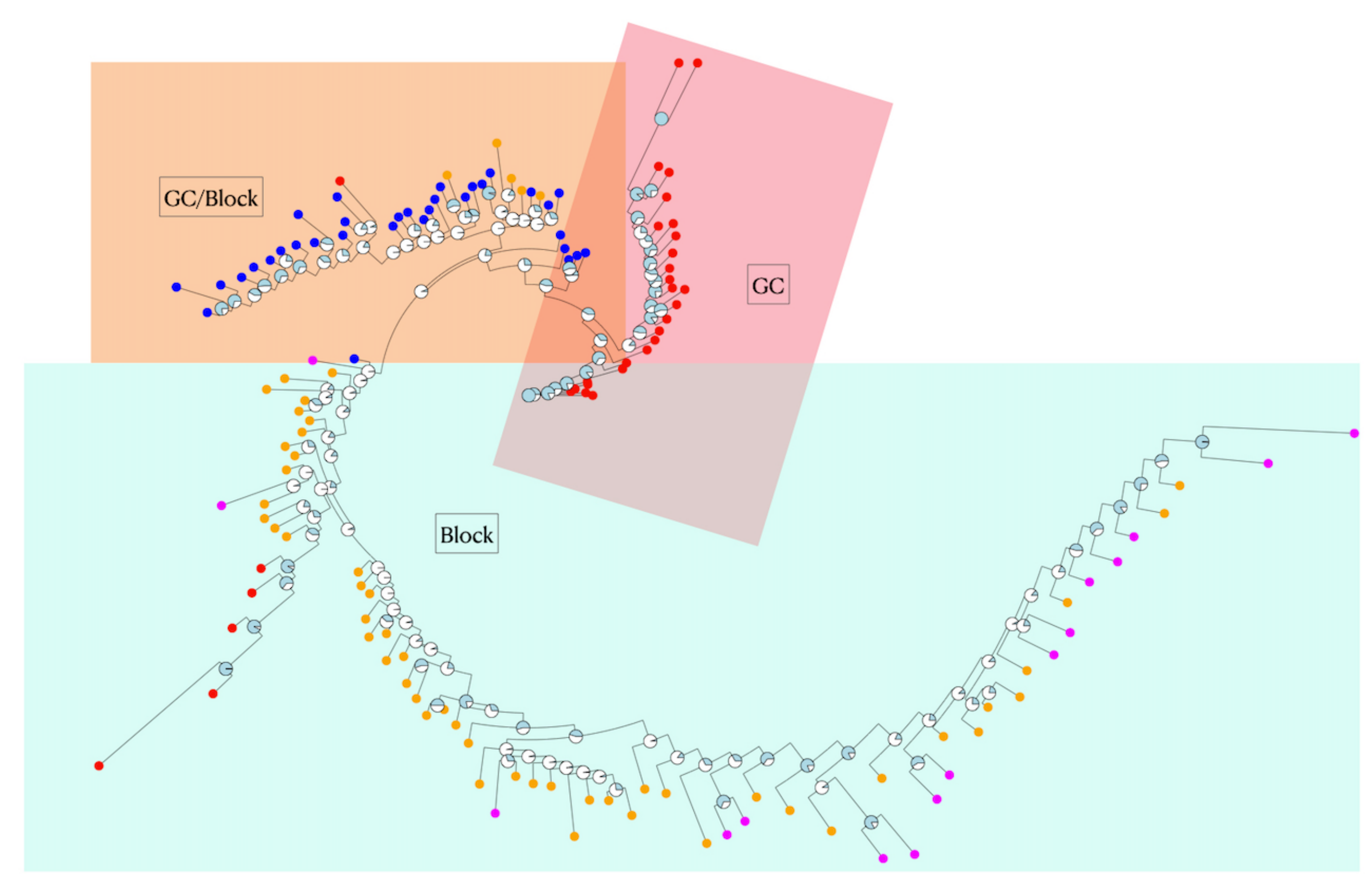}
\caption{Circular Infrared tree. Same as Fig.~\ref{fig:optical_unrooted_tree}}
\label{fig:ir_unrooted_tree}
\end{figure*}

Figure~\ref{fig:ir_unrooted_tree} is the equivalent to Fig.~\ref{fig:optical_unrooted_tree}, and allows us to see the more significant branches and their support.  As in the optical sample, the overall node support in this dataset is 35\%, which is not surprising if the reported abundance uncertainties are comparable to the Optical sample (see more discussions below).

\begin{figure*}[h!]
\centering
\includegraphics[scale=0.15]{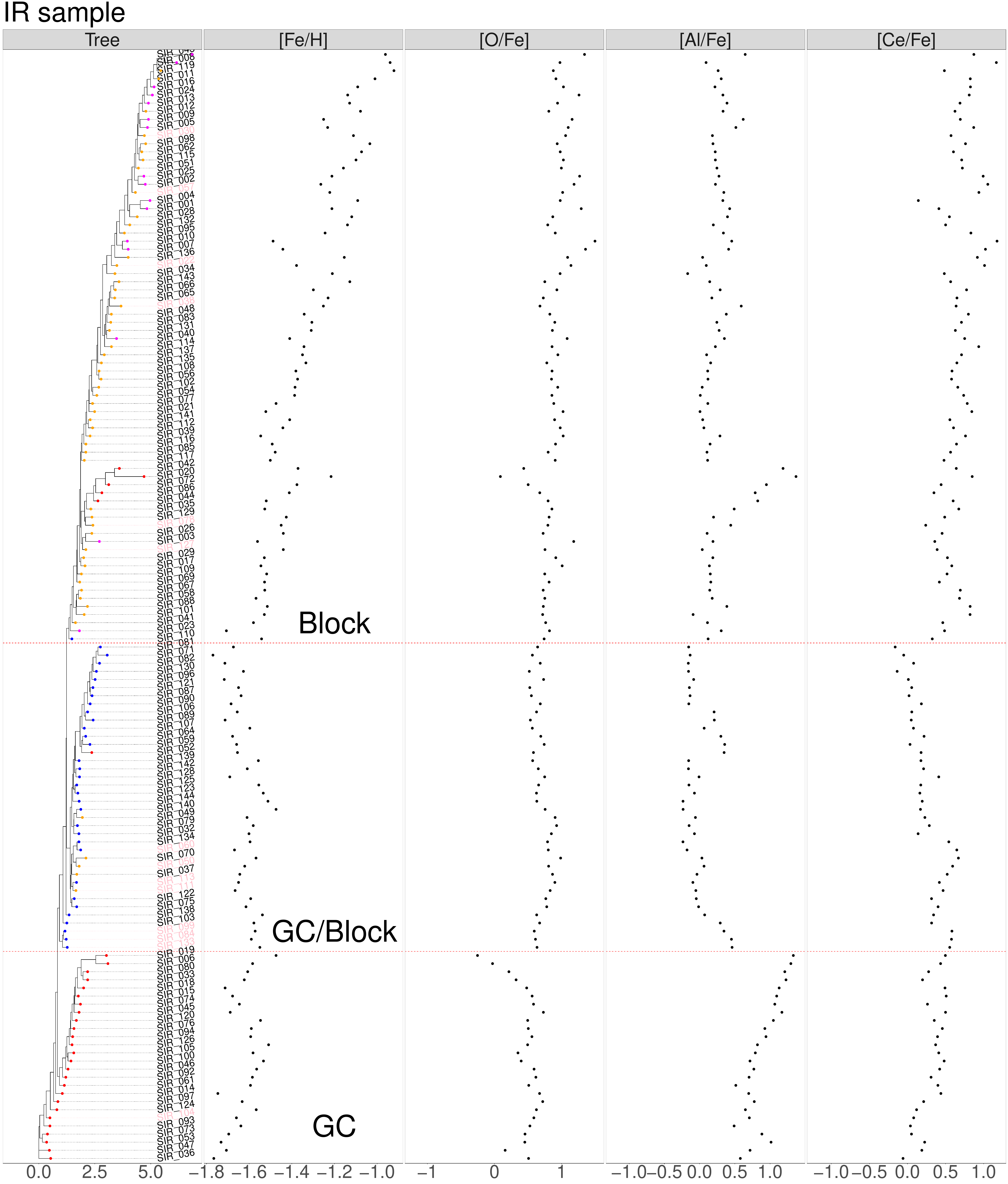}
\caption{Same as Fig.~\ref{fig:tree} but for the Infrared sample. }
\label{fig:hipothesis1_monolithic_ir}
\end{figure*}

Figure~\ref{fig:hipothesis1_monolithic_ir} shows the Infrared tree in the same way as the Optical tree of Fig.~\ref{fig:tree}, which is colored based on the GMM groups illustrated in Fig.~\ref{fig:gmm_structure_ir}. The Infrared tree features more sub-branches than the Optical tree, but only one significant "caterpillar" branch, colored with yellow and magenta. However, we categorize these branches according to the populations identified in the Optical tree, which maps the connections of the stars shared between the two datasets (see sections below). The reason for fewer `caterpillar' long branches in this tree compared to the Optical one might be that this dataset uses fewer abundance ratios, making it harder to divide the tree into populations with different histories. We recall that the stellar population bar plot of the Infrared sample shows a higher proportion of mixed stars compared to the Optical sample and the GMM model prefers four groups instead of three. It is, of course, also possible that the difference in the number of branches is due to selection effects, because we do not have the same stars in both samples, but our analysis of stars in common does not support this idea. 

 Indeed, stars in common with the Optical sample allow us to associate the magenta and yellow stars with {\it Block-Cen}, blue with a mix of Block and GC, and  red  with the {\it GC-Cen stars} (see below).  This is how we can divide the tree of Fig.~\ref{fig:hipothesis1_monolithic_ir} into the upper yellow `caterpillar' branch corresponding to the population {\it Block-Cen}, a middle blue branch being a mix between Block and {\it GC-Cen}, and a lower branch of {\it GC-Cen} stars.  {\it Block-Cen} shows similar chemical properties to the Optical one, namely it has an extended metallicity distribution, with a trend form the lowest metallicity at the root to the highest metallicity at the tip. [O/Fe] and [Al/Fe] decrease with metallicity with comparable values to the Optical branch, and the s-process element ratio [Ce/Fe] increases with metallicity such as [La/Fe]. The scatter in this case is higher, but Ce abundances are uncertain to estimate from the data used in the Infrared sample \citep{All_apo,bawlas}. In general, the trends here have more outliers, such as those red stars at the tip of the lower subbranch. 
 
 The blue branch is composed of both GC and Block stars. Here, no metallicity distribution, as well as no anticorrelation in O-Al is found. Indeed, the blue stars that are also part of the Optical dataset belong to G\_O1-orange, and are thus located at the bottom of the Block branch. This branch may correspond to the first generation of GC stars. The bottom GC branch, colored in red, has no significant metallicity distribution, but shows an anticorrelation, and a [Ce/Fe] positive trend with metallicity. Stars in common between this branch and the Optical suggest they belong to the {\it GC-Cen} population although one star (SIR\_019) is associated with {\it Child-Cen} and it is located at the end of the bottom branch.

As in the Optical case, this tree is not consistent with a single `caterpillar' tree expected from an isolated galaxy evolving like the one analyzed by \cite{DaniTree}.  This provides stronger evidence that \ocen\ is likely the result of more than one history, or/and the result of a complex star formation history.  

\FloatBarrier
\section{Europium abundances and the connection of \ocen\ with GES}\label{sect:europ}

\begin{figure*}[h!]
\centering
\includegraphics[scale=0.1]{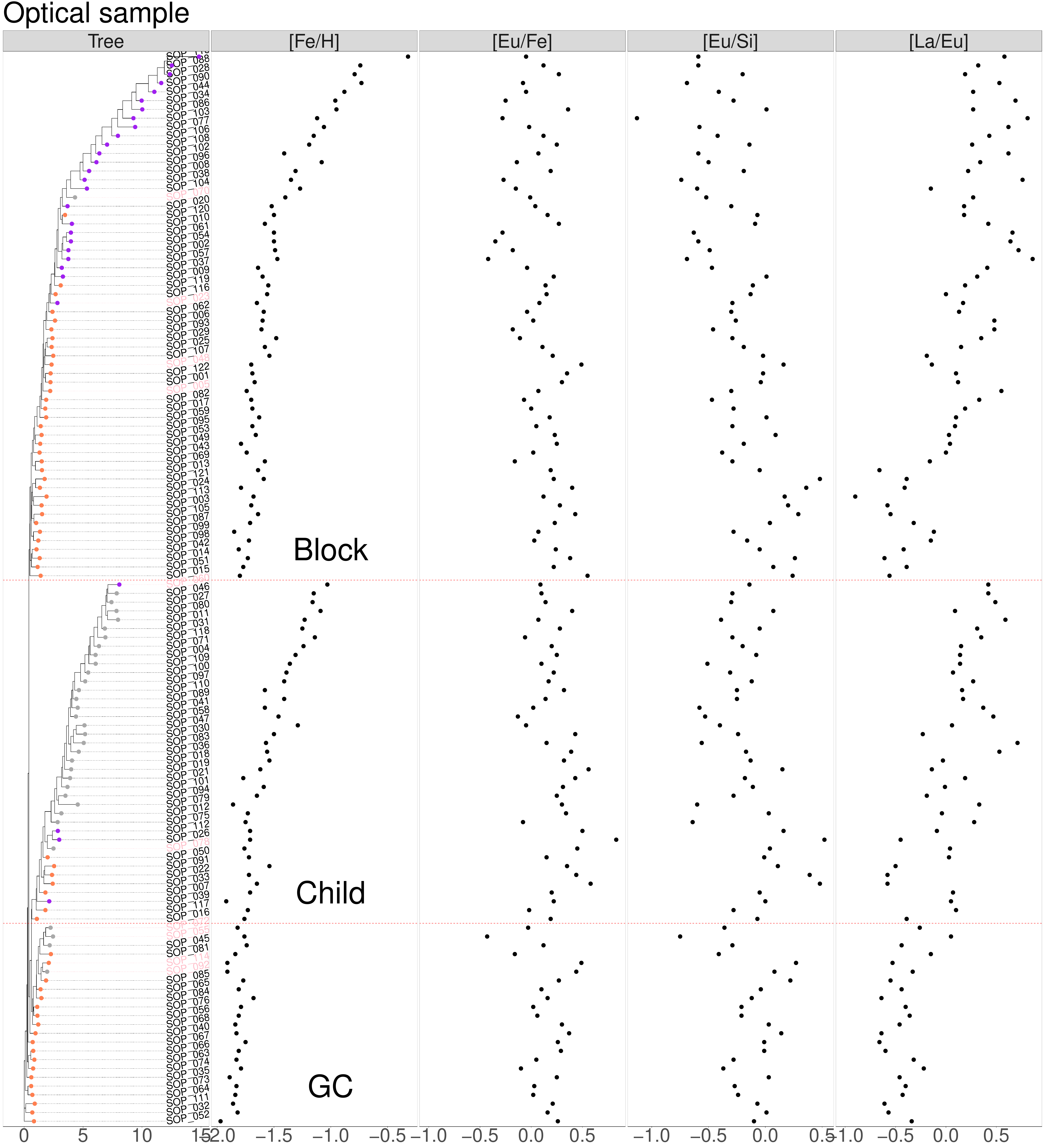}
\caption{Optical tree. Same as Fig.~\ref{fig:tree} but showing the trends of [Eu/Fe], [La/Eu], and [Eu/Si] along the branches}
\label{fig:tree_eu}
\end{figure*}

 In Fig.~\ref{fig:tree_eu} we plot the same data as in Fig.~\ref{fig:tree} but the panels show the trends of [Fe/H], [Eu/Fe],  [Eu/Si], and [La/Eu] along the branches.  In consistency with the scatter plot discussed above, the negative trend of [Eu/Fe] is seen for both, {\it Block-Cen} and {\it Child-Cen}. The trends of [Eu/Fe] however are not so distinct in both branches one another as in the case of [O/Fe] or [Al/Fe] (see Fig.~\ref{fig:tree}), which can be explained by the fact that [Eu/Fe] becomes important only in the third PC of the PCA, unlike O and Al, which matter in the first and the second PC most. The similarity in the [Eu/Fe] abundances in \ocen\ has already been reported by \cite{All_ctio}  {\it Block-Cen} and could mean that r-process elements might not necessarily have been produced inside \ocen\ in large quantities. 

The [La/Eu] and [Eu/Si] trends serve to put in context the possible association of \ocen\ as NSC of the GES. As extensively discussed in the literature, the GES shows significant enhancement of [Eu/Fe] abundances of above 0.5 and even 0.7 dex \citep{aguado21, Matsuno21, carrillo22}, explained by the r-process produced through a neutron star merger (NSM), which can pollute with Eu an entire dwarf galaxy. The ratio of r/s and r/$\alpha$ elements, such as [Eu/La] and [Eu/Si] can be used to study the importance and the time delay of the r-process pollution, because of the different lifetimes of the different progenitors. In fact, \cite{Matsuno21} found that GES has a flat trend of [La/Eu] with scatter, in agreement with the trends of [Ba/Eu] of \cite{aguado21}, consistent with Eu being produced by a NSM event. Because GES has low abundances of $\alpha-$elements compared to the in-situ halo, \cite{Monty24} uses [Eu/Si] to distinguish between accreted and in-situ stars, showing that accreted stars have significantly higher [Eu/Si] compared to in-situ stars. This includes the globular clusters.

Figure~\ref{fig:tree_eu} shows that all branches and GMM groups have comparable distributions in [Eu/Fe], and have values that are below the reported ones for GES. The {\it Block-Cen} and the {\it Child-Cen} branches have increasing [La/Eu] trends with metallicity, suggesting that s-process is important in these populations. [Eu/Si] is low in {Block-Cen} and has a decreasing trend with metallicity, while for Child-Cen we can not identify a clear trend. The {\it GC-Cen} has solar-values for [Eu/Si], and no trend. These abundance trends might rule out the chemical connection of \ocen\ with the GES, but only if we believe that the NSC of the GES has to follow the same chemical trends as the rest of the galaxy, which is not established so far. Indeed, if there was a time-delay to pollute the GES with r-process abundances from a NSM, then it is possible that {\it Block-Cen} and {\it GC-Cen} formed before this event happened. We recall that $\alpha-$capture element abundances of this population are higher than the typical GES ones. The fact that the Eu abundances do not increase in {\it Child-Cen} might also suggest that no other NSM event happened inside \ocen.

\end{appendix}

\end{document}